\documentclass[reprint,amsmath,amssymb,aps,prmaterials, superscriptaddress,floatfix]{revtex4-1}
\usepackage[latin9]{inputenc}
\usepackage{amsmath}
\usepackage{graphicx}
\usepackage{braket}
\usepackage{hyperref}

\makeatletter

\providecommand{\tabularnewline}{\\}

\newcommand*{\chem}[1]{\ensuremath{\mathrm{#1}}}

\makeatother

\begin{document}
\title{Contrasting Ferromagnetism in Pyrite \chem{FeS_2} Induced by Chemical
Doping versus Electrostatic Gating}
\author{Ezra Day-Roberts}
\affiliation{School of Physics and Astronomy, University of Minnesota, Minneapolis,
MN 55455, USA}
\author{Turan Birol}
\affiliation{Department of Chemical Engineering and Materials Science, University
of Minnesota, Minneapolis, Minnesota 55455, USA}
\author{Rafael M. Fernandes}
\affiliation{School of Physics and Astronomy, University of Minnesota, Minneapolis,
MN 55455, USA}
\begin{abstract}
Recent advances in electrostatic gating provide a novel way to modify
the carrier concentration in materials via electrostatic means instead of chemical doping, thus
minimizing the impurity scattering. Here, we use first-principles Density Functional Theory combined with a tight-binding
approach to compare and contrast the effects of electrostatic gating
and Co chemical doping on the ferromagnetic transition of \chem{FeS_2}, 
a transition metal disulfide with the pyrite structure.
Using tight-binding parameters obtained from maximally-localized Wannier
functions, we calculate the magnetic susceptibility across a wide doping range. We find that electrostatic
gating requires a higher electron concentration than the equivalent
in Co doping to induce ferromagnetism via a Stoner-like mechanism.
We attribute this behavior to the formation of a narrow Co band near
the bottom of the conduction band under chemical doping, which is
absent in the electrostatic gating case. Our results reveal that the effects
of electrostatic gating go beyond a simple rigid band shift, and highlight
the importance of the changes in the crystal structure promoted by
gating.
\end{abstract}
\maketitle

\section{Introduction}

Transition metal disulfides, $\mathrm{(TM)S}_{2}$, with the pyrite structure 
host a wide variety of electronic ground states \cite{Ogawa1976}.
Varying the transition metal TM tunes the band filling over a wide
range, from a $3d^{5}$ electronic configuration in the case of \chem{MnS_2}
to a $3d^{10}$ configuration for \chem{ZnS_2}. As the carrier concentration
changes, a rich landscape of electronic states emerges, including:
an antiferromagnetic insulator (\chem{MnS_2})\cite{Chattopadhyay_1991,LIN1968687,Kimber_2015}, a semiconductor (\chem{FeS_2}) \cite{PhysRevB.48.15781,leighton_composition_2007},
a ferromagnetic metal (\chem{CoS_2}) \cite{PhysRevB.48.15781,leighton_composition_2007}, an antiferromagnetic Mott insulator
(\chem{NiS_2}) \cite{PhysRevB.68.094409,Schuster2012,Yano2016}, a superconductor (\chem{CuS_2}) \cite{BITHER1966533,PhysRevB.65.155104}, and another semiconductor
(\chem{ZnS_2}) \cite{Bullett_1982}. Tuning continuously
across these phases would provide a unique avenue to elucidate the
interplay between different electronic orders. 
%
%
While it is possible to use chemical doping to move across most of the transition metal disulfides' phase diagram,  this approach introduces disorder and local inhomogeneity, which complicates the theoretical picture \cite{goldman_electrostatic_2014}. 

Electrostatic gating offers a promising alternative to chemical doping
as a means to tune the carrier concentration, while avoiding the steric
and chemical (electronegativity, etc.) effects associated with the addition of dopants.
While the effects achievable using a conventional gate dielectric 
are often limited, novel gating approaches such 
as using a polar oxide or ferroelectric gating are quite promising  \cite{Burton2011, Liu2018}. Also exciting are the recent advances in electrostatic gating with ionic liquids or gels, which provide
access to much higher electron concentrations than those attainable
by dielectric-based gating \cite{goldman_electrostatic_2014, Zhang1523686, Song_2016}, opening new avenues to explore different regions of electronic phase 
diagrams \cite{Scherwitzl2010,Ueno2008,Jeong2013,Bisri2017}, including wide regions of the disulfide pyrite electron-density phase diagram. 
Indeed, in dielectric-based gating devices, breakdown voltages restrict the added carrier densities
to values $<10^{12-13}\text{cm}^{-2}$ \cite{goldman_electrostatic_2014}. In contrast,
the ability to achieve carrier concentrations of up to $8\times 10^{14}\text{cm}^{-2}$ \cite{10.1002/adfm.200801633} via electrolyte gating has been widely employed to study a variety of
phenomena in oxides, such as the structural
transformation in \chem{VO_2} \cite{Jeong1013,10.1063/1.4861901,10.1021/acs.nanolett.6b01756}, the metal-insulator
transition in \chem{SrRuO_3} \cite{Yi2014,10.1063/1.4899145}, and  superconductor-insulator
transitions in multiple materials \cite{Ueno2008,PhysRevB.92.165304,Bollinger2011,PhysRevLett.107.027001,Ye1193}. These studies, however,
revealed an important issue associated with electrolyte gating:
often, electrochemical effects beyond simple electrostatics are at play \cite{Zhang1523686, Petach2014,Bubel2015,Leighton2019}. 
For example, in La$_{0.5}$Sr$_{0.5}$CoO$_{3-\delta}$, oxygen vacancies are formed under positive gating 
voltages \cite{PhysRevMaterials.1.071403}. These vacancies, which are formed in response to gating, 
enhance the sensitivity of the electronic structure to gating. However, 
they also introduce a significant degree
of irreversibility. While this irreversibility can be undesirable for certain applications, 
many attempts have been made to take
advantage of these electrochemical effects for many applications 
as well \cite{Chen2009, Yuan2010, Kay2012, Ge2015, PhysRevMaterials.1.071403}. 
%
%
In contrast to oxides, strong sulfur-sulfur bonding in the pyrite structure 
\cite{ramesha_experimental_2004} makes the formation energy of single sulfur vacancies 
prohibitely high \cite{Sun2011} while multi-vacancy defect complexes dominate the electrochemical 
response \cite{Gagliardi}. How these defect complexes diffuse and determine the 
electrochemical response in pyrites is far from clear. 

Among the pyrite transition metal disulfides, \chem{FeS_2} has attracted interest both as a
potential photovoltaic material, characterized by a high optical absorption,
a low toxicity and a low cost to manufacture \cite{ALTERMATT2002181,10.1021/nl202902z,ALHARBI20112753}, and as a metallic ferromagnet when doped with cobalt \cite{10.1002/pssb.200666821,leighton_composition_2007}.
\chem{FeS_2} is often unintentionally doped, and a great amount of work has been performed on the nature of native dopants  \cite{Sun2011, Voigt2019} as well as the role of surface vs. bulk conduction \cite{Limpinsel2014, Liang2014}. 
While the ``doping puzzle'' about the nature of native dopants in single crystals vs. films of FeS$_2$ seems to be resolved \cite{Zhang2017, Gagliardi}, there are several open questions about the electronic properties
of \chem{FeS_2} that remain unsettled, such as the impact of the
conducting surface states \cite{walter_surface_2017}, the nature
of the ferromagnetic transition in the doped compounds \cite{guo_discovery_2008},
and the role of Co doping in inducing ferromagnetism even at very
small doping concentrations \cite{PhysRevB.81.144423}.

To shed new light on some of these issues, in this paper we perform
a first-principles study of electrostatically gated \chem{FeS_2}
and \chem{CoS_2} with pyrite structure, systematically comparing their electronic and magnetic
properties with those of chemically doped \chem{Fe_{1-x}Co_xS_2}.
To model electrostatic
gating, we go beyond the rigid-band shift paradigm and account for
changes in the band structure and in the crystal structure arising
from the change in the carrier concentration \cite{kotiuga2019highdensity}. By computing the magnetization,
we find that ferromagnetism appears for a smaller added carrier concentration
in the case of chemical doping as compared to electrostatic
gating. We attribute this behavior to the different energy ranges of the
wide sulfur anti-bonding band in the two cases, as well as to the
existence of a narrow Co band near the bottom of the conduction band
in the case of chemical doping.

By comparing the carrier-concentration evolution of the magnetization
with that of the density of states, we propose that the ferromagnetism
is promoted by the Stoner mechanism. This naturally accounts for the
sensitivity of the ferromagnetism to the changes in the band structure
caused by chemical doping and electrostatic gating. We go beyond the
first-principles analysis by computing the Lindhard function from
a multi-orbital tight-binding model derived from the maximally localized
Wannier functions. We find that the non-interacting magnetic susceptibility
is peaked at the $\Gamma$ point of the Brillouin zone, confirming
the Stoner-character of the ferromagnetic instability and ruling out
finite wave-vector magnetic states.

The paper is organized as follows: In section \ref{sec:Methods} we
summarize our methods. In section \ref{sec:First-Principles-Results}
we present our first-principles results on the magnetic, electronic,
and crystalline structures of the chemically doped and electrostatically
gated cases. In section \ref{sec:Tight-Binding-Model} we fit a tight
binding model to our first-principles results to calculate the non-interacting
magnetic susceptibility. We conclude with a summary of our main results
in section \ref{sec:Conclusions-and-Summary}.

\section{Methods}
\label{sec:Methods} 

DFT+U calculations were done using the VASP implementation of
the projector-augmented-wave (PAW) approach \cite{VASP,VASP-PAW}.
The exchange-correlation functional was approximated using the PBEsol set 
generalized gradient approximation (GGA), which is developed
for accuracy in crystal structure relaxations \cite{PBEsol}. To correct the underestimation 
of on-site interactions between electrons, DFT+U approach was used  \cite{Dudarev1998}.
A value of $U=5$~eV
was selected as a compromise to achieve good agreement with the experimental
lattice constant and sulfur-sulfur distance for both \chem{FeS_2}
and \chem{CoS_2} (see Appendix).
For \chem{FeS_2} alone, a lower value of approximately $2$~eV is
optimal, in agreement with previous works \cite{hu_first-principles_2012}.
For \chem{CoS_2} alone, a much larger value of $U$ is preferred,
because the lattice constant is underestimated and the sulfur-sulfur
distance is overestimated for all values below $7$~eV. The $U$
value of $5$~eV gives an error in each lattice constant of less than
$1\%$ and an error in the sulfur-sulfur distance of about $2.5\%$.
A $\Gamma$-centered $k$-point grid of $8\times8\times8$ was used
for structural calculations along with a plane wave cutoff of $500$~eV.

Structural parameters for chemically doped and electrostatically gated
systems are determined by performing structural relaxations. These
calculations allow spin-polarization to reflect the presence of local
moments, which is important for obtaining realistic crystal structures
in DFT. Undoped \chem{FeS_2} is found to not be spin polarized in
its ground state while for all other chemical doping levels the ground
state is found to be spin polarized, consistent with previous reports \cite{mazin_robust_2000, feng_structural_2018}. A tight-binding model is constructed by calculating
the Maximally Localized Wannier Functions by employing the wannier90
package \cite{wannier90}. The wannierization calculation is done in
the non-spin polarized state with the same value of $U$, since we
are concerned with the emergence of the magnetic instability and not
with the behavior of the materials in their ferromagnetic state.

In order to compare the effects of electrostatic gating with
chemical doping we performed two sets of calculations. To simulate
electrostatic gating (EG) we consider undoped \chem{FeS_2} (or \chem{CoS_2})
and vary the total number of electrons in the unit cell by adding
electrons (or holes) \footnote{As is standard in similar DFT calculations, a 
homogeneous background charge is also added to ensure charge neutrality of the unit cell.}. 
The highest level of EG we considered was 1 removed electron or 0.5 added electrons per transition metal atom. This is almost an
order of magnitude larger than the experimentally achievable values,
and leads to very large changes in the crystal structure. Thus, the
results presented for the highest levels of EG serve just to illustrate
trends for comparison with chemical doping. To simulate chemical doping
(CD), we replaced one, two, or three of the \chem{Fe} ions in the
unit cell with \chem{Co} ions, corresponding to $x=0.25,0.50,\text{ and }0.75$
in \chem{Fe_{1-x}Co_xS_2}. For each different carrier concentration
in EG we fully relaxed both the ionic positions and the lattice vectors.
For the CD configurations, the dopants break the symmetry of the crystal
structure, so cell shape distortions away from cubic are in principle
permitted by symmetry. Since the average structure with disorder has
cubic symmetry, such distortions were not allowed in our calculations.
This was achieved by iteratively relaxing cell size and ionic positions
separately until convergence was obtained.

\begin{figure}
\includegraphics[width=0.45\textwidth]{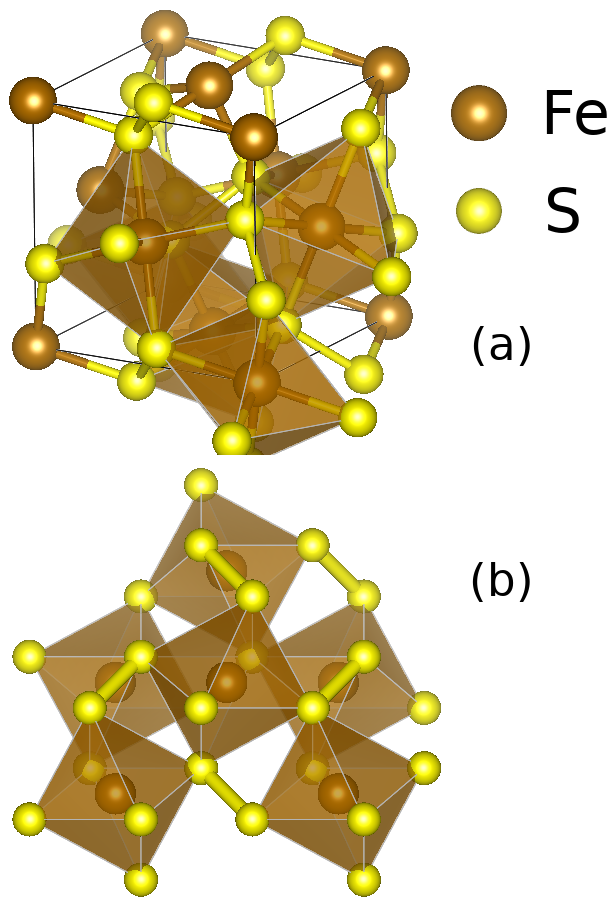} 
\caption{(a) The simple-cubic primitive unit cell of \chem{FeS_2}. The transition
metal atoms occupy the corners and the face centers of this cell.
Each transition metal atom is in the center of a sulfur octahedron.
(b) The transition metal octahedra are corner sharing in the pyrite
structure. In addition, every sulfur is part of a dimer connecting
neighboring octahedra.}
\label{fig:cryst}
\end{figure}

The entire conduction band manifold that consists of 2 $e_{g}$ orbitals
and 1 sulfur anti-bonding orbital per FeS$_{2}$ formula unit was
used for wannierization. This manifold is isolated from other bands 
so no disentanglement was necessary \cite{wannier_entangled}.
The tight binding models we obtained from the wannierization procedure
reproduce the DFT band structure extremely well (see the Appendix for details),
but this requires using a very large number of hopping parameters.
As a result, we do not report our hopping parameters.

\section{First-Principle Results}

\label{sec:First-Principles-Results}

\subsection{Magnetization}

\begin{table}
\begin{tabular}{l|c|c}
 & Lattice Constant (\AA) & Internal parameter (u)\tabularnewline
\hline 
\chem{FeS_2} Exp & 5.428 & 0.385\tabularnewline
\chem{FeS_2} Theory & 5.421 & 0.387\tabularnewline
\hline 
\hline 
\chem{CoS_2} Exp & 5.535 & 0.395\tabularnewline
\chem{CoS_2} Theory & 5.510 & 0.391\tabularnewline
\end{tabular}\caption{Experimental crystal structure parameters for pure \chem{FeS_2} and
\chem{CoS_2} compared with our calculations \cite{Finklea:a12931,10.1002/zaac.19382390110}.}
\label{tab:params}
\end{table}

The pyrite structure has a simple cubic cell, consisting of a face-centered
lattice of transition metal atoms each surrounded by a distorted sulfur
octahedron (see Fig. \ref{fig:cryst}). The sulfur atoms form covalently
bonded dimers, with the center point of the dimers forming another
FCC lattice shifted from the transition metal lattice by half a lattice
vector \cite{Nowack:bx0513}. Because the sulfur atoms share two electrons
in these dimers, the sulfur charge state is $-1$. This results in
a total charge of $-2$ per dimer, and hence the iron atoms have an
$\mathrm{Fe}^{2+}$ valence. This is in contrast to oxides and most other
transition metal dichalcogenides where the chalcogens have a $-2$ charge,
implying an $\mathrm{Fe}^{4+}$ valence \cite{Streltsov2017}. In \chem{FeS_2}, the dimer
anti-bonding states are unoccupied and overlap with the empty \chem{Fe}
$e_{g}$ bands. These dimers are thus an important ingredient of the
electronic structure. The sulfur-sulfur distance controls the energy 
of the sulfur anti-bonding bands that make up the bottom of the conduction
band in \chem{FeS_2}. There is only one internal crystallographic parameter,
$u$, which controls the sulfur atoms' positions at $(u,u,u)$ and
at the symmetry-equivalent positions. This parameter controls both
the distortion of the octahedra and the relative sulfur-sulfur and
transition metal-sulfur distances. Table \ref{tab:params} lists the
previously reported experimental lattice constant and internal parameter \cite{Finklea:a12931,10.1002/zaac.19382390110}
, comparing them to the relaxed values found in
this work. The discrepencies come from our choice to use a single value of $U$ for both FeS$_2$ and CoS$_2$.

\begin{figure}
\includegraphics[width=0.5\textwidth]{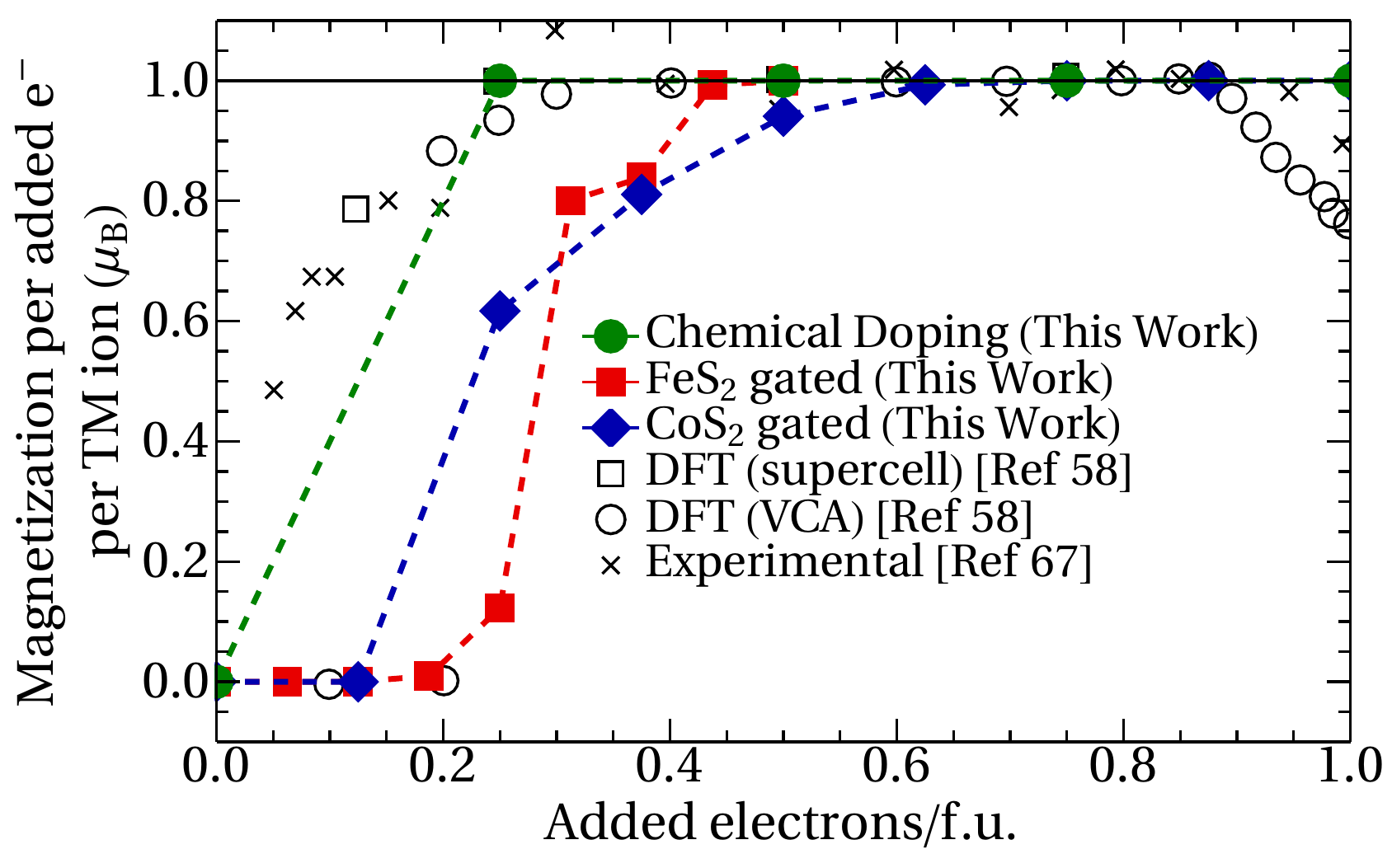}
\caption{Magnetization per added free electron as function of the number of
added electrons. Our DFT+U results are compared with previous first-principles
and experimental results on doped \chem{Fe_{1-x}Co_xS_2} \cite{mazin_robust_2000}.
Note that electrostatically gated systems do not achieve 100\% spin
polarization until approximatelly $0.4$ electrons are added to \chem{FeS_2}.
Furthermore, electrostatic gating requires a higher added electron
concentration to achieve ferromagnetism as compared to chemical doping.}
\label{fig:magnetization_comparison}
\end{figure}

The magnetic transition that takes place on going from \chem{FeS_2}
to \chem{CoS_2} allows access to a large range of spin polarizations
\cite{leighton_composition_2007}, making this possibly the best studied
transition in the pyrite disulfides family. In early experiments, ferromagnetism
was found already at very low doping levels of less than $x=0.01$
in \chem{Fe_{1-x}Co_xS_2} \cite{jarrett_evidence_1968}, an observation
that has been confirmed by many later experiments \cite{guo_discovery_2008, leighton_composition_2007, ramesha_experimental_2004}
(see experimental points in figure \ref{fig:magnetization_comparison}).
Magnetization measurements show that \chem{Fe_{1-x}Co_xS_2} is a
nearly perfect half-metal across a large range of doping concentrations
($x\approx0.1-0.6$).

\begin{figure}
\includegraphics[width=0.4\textwidth]{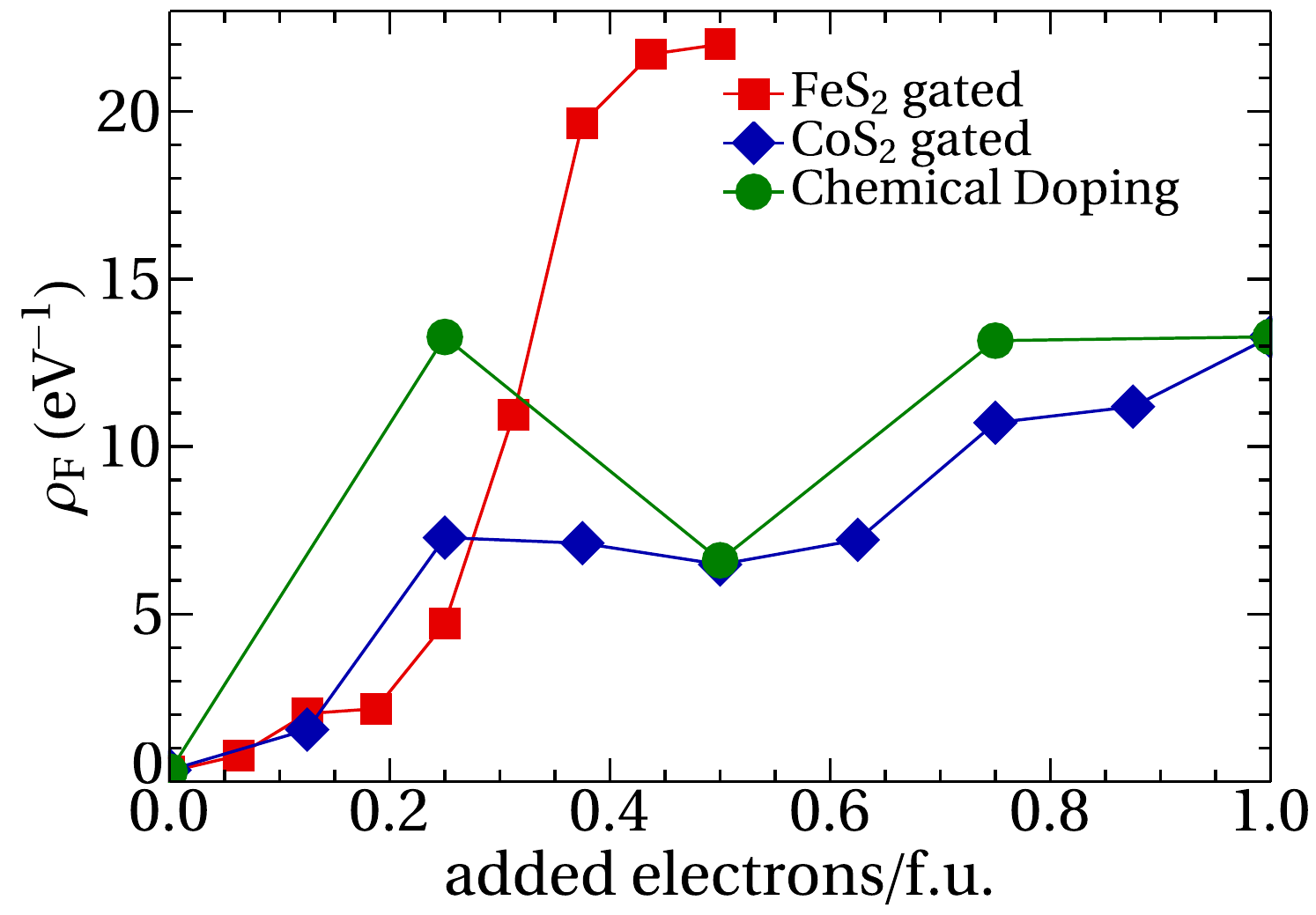} 
\caption{Calculated density of states at the Fermi level $\rho_{F}$as function
of the number of added electrons for electrostatically gated \chem{FeS_2}
(red squares) and \chem{CoS_2} (blue diamonds), and chemically doped
\chem{Fe_{1-x}Co_xS_2} (green circles).}
\label{fig:gamma} 
\end{figure}

\begin{figure*}
\includegraphics[width=0.9\textwidth]{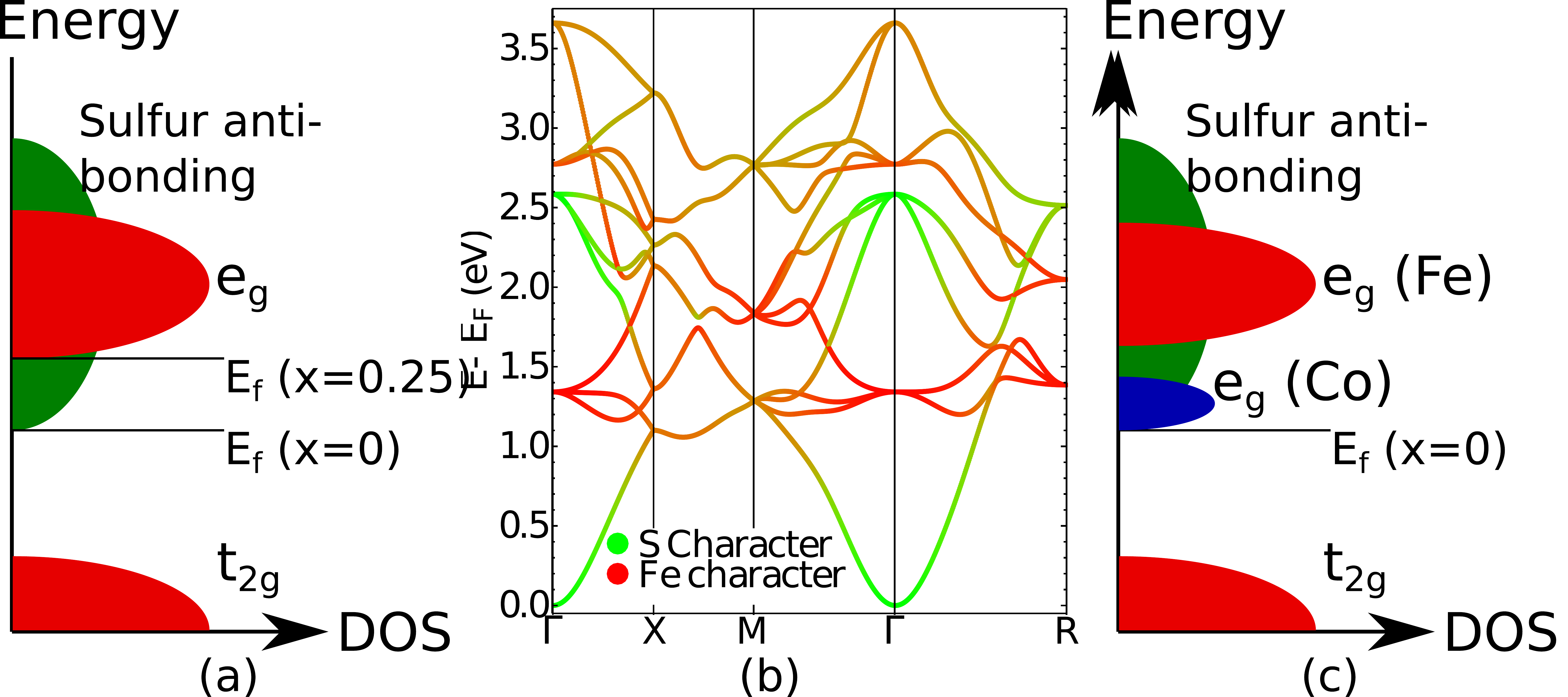}
\caption{\label{fig:schematicDOS} (a) Schematic representation of the DOS
for electrostatic gated \chem{FeS_2}. The red (green) bands are
iron (sulfur) bands. (b) DFT band structure of \chem{FeS_2}, with
red denoting greater iron character and green, greater sulfur character.
(c) Schematic representation of the DOS for chemically doped \chem{Fe_{1-x}Co_xS_2}.
A cobalt $d$-band (blue) emerges at the bottom of the wide sulfur
band.}
\end{figure*}

First-principles calculations predict ferromagnetism at a larger value
$x\approx0.10-0.15$, and a half-metal with 100\% spin polarization
emerging at and above $x\approx0.20-0.25$ \cite{mazin_robust_2000,feng_structural_2018}.
These results for the magnetization (open square and circles), combined
with our own DFT+U results for chemically doped \chem{Fe_{1-x}Co_xS_2}
(green circles), are shown in Fig. \ref{fig:magnetization_comparison}.
In agreement with previous results, we find a ferromagnetic transition
occuring for $0<x<0.25$. In contrast, in the case of electrostatically
gated \chem{FeS_2} (red squares), ferromagnetism onsets only at larger
carrier concentrations, equivalent to $x\approx0.20-0.30$, with half-metallicity
appearing only at $x\approx0.40$. Conversely, starting from \chem{CoS_2}
and adding holes (blue diamonds), half-metallicity starts disappearing
around $1-x\approx0.4$. The reasons for these differences will be
explored in the next subsections, were we contrast the band structure
and crystal structure parameters in the cases of chemical doping and
electrostatic gating.

\subsection{Density of States}

To shed light on the origin of the ferromagnetic state, we plot in
Fig. \ref{fig:gamma} the DOS at the Fermi level, $\rho_{F}$, as
function of the added carrier concentration. Comparison with the behavior
of the magnetization in Fig. \ref{fig:magnetization_comparison} suggests
that a Stoner mechanism is likely at play \cite{mazin_robust_2000}. Indeed, at low carrier concentrations,
the CD material has a higher DOS at the Fermi level than the EG material,
consistent with the fact that the former is ferromagnetic at low doping
levels. Similarly, the DOS of the EG materials show a significant
increase around $x\approx0.25$, which coincides with the onset of
ferromagnetism in Fig. \ref{fig:magnetization_comparison}.

\begin{figure}
\includegraphics[width=0.23\textwidth]{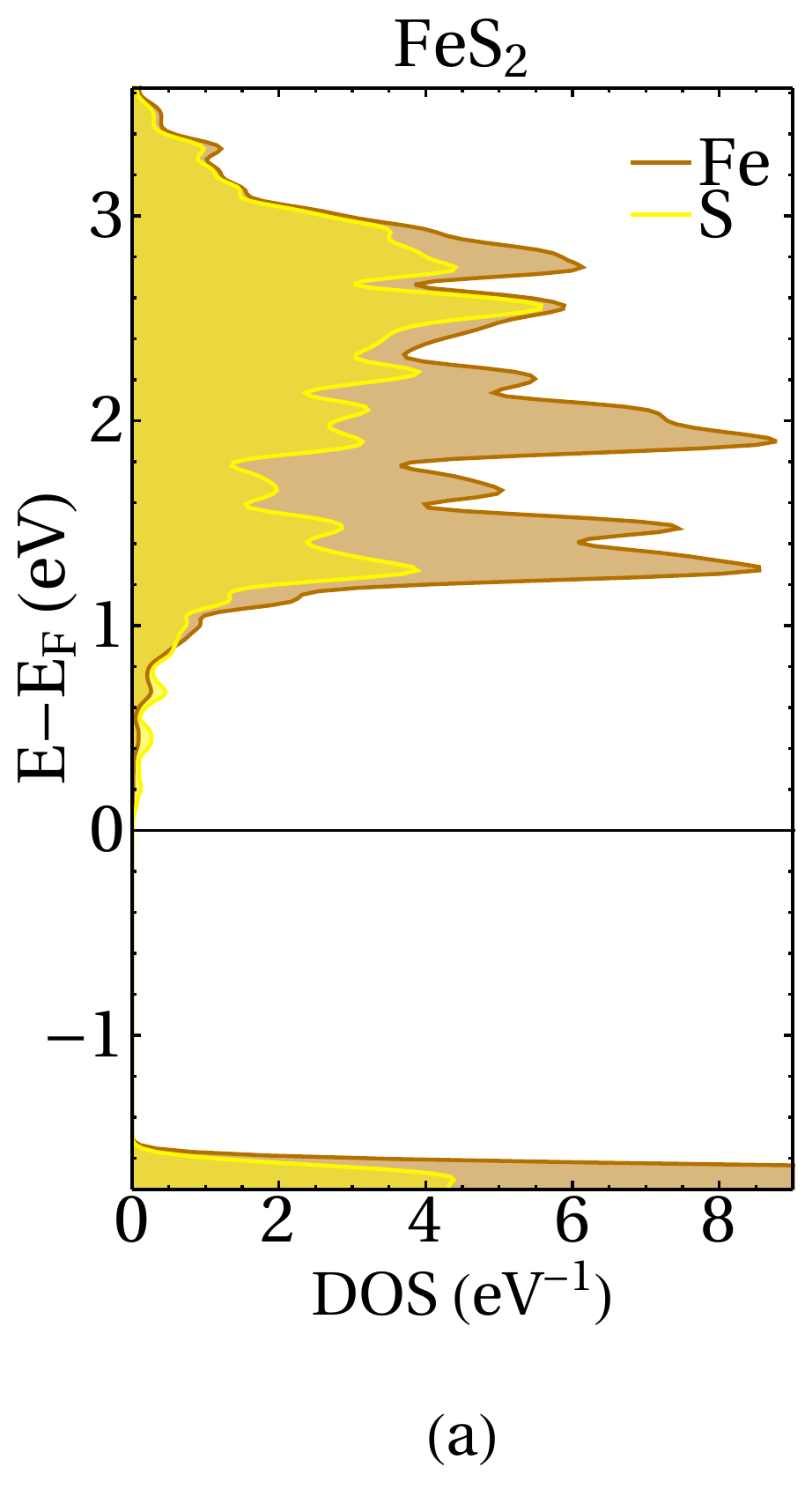}
\includegraphics[width=0.23\textwidth]{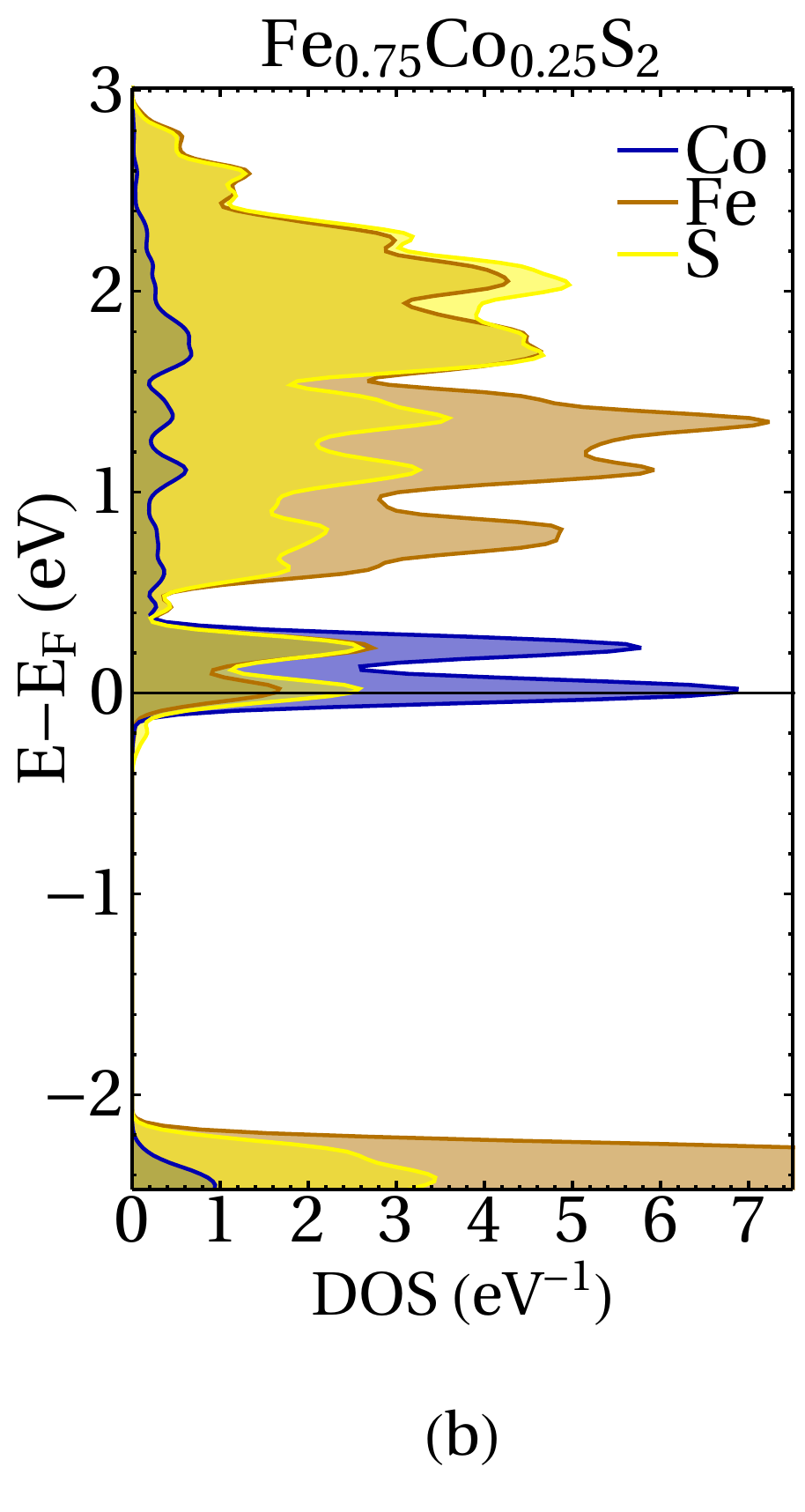}
\label{fig:calcChemDOS}
 \caption{\label{fig:calcDOS} Calculated densities of states for (a) pure \chem{FeS_2} and (b) chemically doped \chem{Fe_{0.75}Co_{0.25}S_2}. The character of the bands are colored according to the legends. Note the prominent peak originating from the Co orbitals at the bottom of the conduction band in the doped case.}
\end{figure}

The key difference between EG and CD compounds is which bands are
being filled. Figure \ref{fig:schematicDOS}(a) shows a schematic
representation of the density of states for EG \chem{FeS_2}, whereas
the calculated DOS is shown in Fig. \ref{fig:calcDOS}(a). The valence
band consists of fully occupied $t_{2g}$ orbitals, and the conduction
band consists of unoccupied $e_{g}$ states surrounded by a wide sulfur
$p$-band \cite{Eyert1998}. This wide band has sulfur-sulfur antibonding character,
as shown in Fig. \ref{fig:schematicDOS}(b) \cite{Folkerts1987}. Gating affects the relative bandwidth of the sulfur bands, which decreases for increasing $x$. Introducing electrons to \chem{FeS_2} initially
fills this sulfur band, which has a low density of states. \chem{Fe}
$e_{g}$ states start being occupied only after around $0.25$ electrons
per iron are added. Once the $e_{g}$ band starts being filled, the
DOS increases significantly, and ferromagnetism emerges. This qualitative
picture also applies to EG \chem{CoS_2}, although it has a narrower
sulfur bandwidth as compared to EG \chem{FeS_2}.

The DOS evolution with carrier concentration is rather different in
the case of CD \chem{Fe_{1-x}Co_xS_2}. The reason is because, in
the $2+$ valence state, iron is slightly more electronegative than
cobalt (1.390 vs 1.377 in the scale defined by Yuan et al. \cite{electronegativities}),
which means that the same electronic orbitals will be at lower energies
in cobalt relative to iron. As a result, occupied \chem{Co} $e_{g}$
states are lower than the unoccupied $e_{g}$ states of \chem{Fe}.
These states, which have a large DOS, make up the lower edge of the
conduction band as illustrated schematically in Fig. \ref{fig:schematicDOS}(c)
and shown quantitatively in Fig. \ref{fig:calcDOS}(b). Thus, in contrast
to the electrostatically gated case, where the added electrons start
by occupying low DOS sulfur states, the extra electrons in \chem{Fe_{1-x}Co_xS_2} occupy high DOS cobalt $e_{g}$ states immediately.
This is related to the appearance of ferromagnetism at a much lower
added carrier concentration.

\subsection{Crystal Structure}

\begin{figure}
\includegraphics[width=0.42\textwidth]{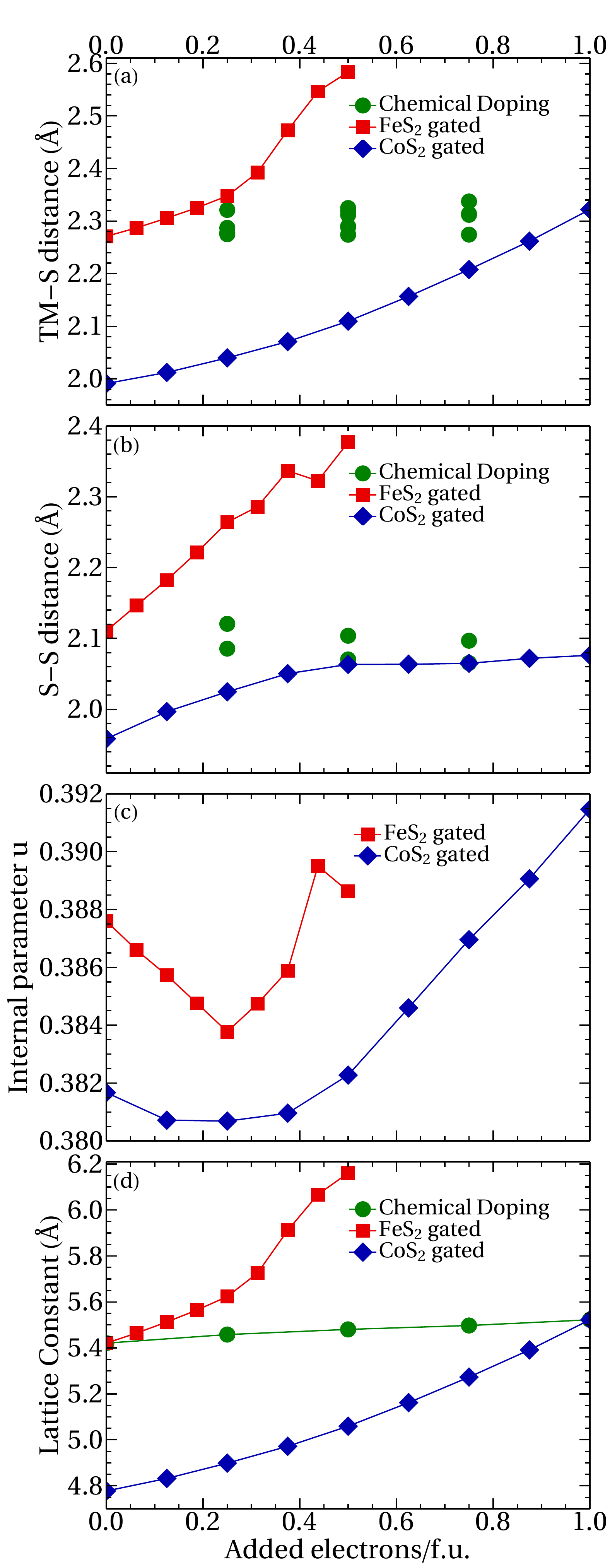}
\caption{Plots of several structural parameters as a function of added electrons
(per formula unit) for electrostatically-gated \chem{FeS_2}
(red), electrostatically-gated \chem{CoS_2} (blue, in which case
the added carriers are actually holes), and chemically doped \chem{Fe_{1-x}Co_xS_2}
(green). The panels display (a) the sulfur-sulfur distance, (b) the
transition metal-sulfur distance, (c) the internal sulfur parameter
$u$, and (d) the lattice constant. 
Chemical doping reduces the symmetry and leads to multiple different TM--S and S--S distances, 
which are shown as separate datapoints in panels (a) and (b)
}	
\label{fig:dists}
\end{figure}

The changes in the electronic structure discussed above are also accompanied
by changes in the crystal structure, highlighting the importance of
effects beyond a simple rigid-band shift in the case of electrostatically
gated compounds. Figure \ref{fig:dists} shows the evolution of the
crystal structure with increasing electron count, contrasting the
cases of electrostatic gating (red and blue curves) and chemical doping
(green curves). The former is modeled either as electrons added to
the \chem{FeS_2} structure (red) or as holes added to the \chem{CoS_2}
compound (blue). There are noticeable changes in the trends of multiple
structural parameters near $0.25\text{\textendash}0.30$ added
electrons per formula unit (f.u.). While some changes
might seem unphysically large, we emphasize that large values of added
carriers are not experimentally feasible via electrostatic gating,
and are only included here to illustrate the trends.

These effects are mainly driven by the Fermi level entering the $e_{g}$
bands at this doping, as discussed in the previous subsection. For
less than $0.25$ added electrons per f.u., the states that are being
filled have sulfur antibonding character, which causes the sulfur-sulfur
distance to increase (panel (b)). Once $0.25$ electrons per f.u.
are added, the $e_{g}$ bands begin filling, which is reflected in
the sharp upturn in the transition metal-sulfur distance in the case
of \chem{FeS_2} (panel (a)). In the case of \chem{CoS_2}, there
is a much less steep change, although an increase is also observed.
At the same time, since there are still some sulfur-sulfur antibonding states
at the Fermi level, the lattice constant increases at a faster rate
(panel (d)) to compensate for the effects of the internal parameter
$u$ (panel (c)). This behavior of the lattice constant under electrostatic
gating strongly deviates from a linear interpolation that would be expected from
Vegard's law \cite{Vegard1921}, which is well followed under chemical
doping. Indeed, adding $0.25$ electrons per \chem{Fe} increases
the lattice constant by more than $4\%$, whereas $25\%$ \chem{Co}
doping only changes the lattice constant by $\approx.5\%$.

These two effects impact the evolution of the internal parameter $u$,
shown in panel (c). This parameter and the lattice constant $a$ are
related to the sulfur-sulfur and metal-sulfur distances as $d_{\mathrm{S-S}}=a\sqrt{3}(1-2u)$
and $d_{\mathrm{TM-S}}=a\sqrt{\frac{1}{2}-2u+3u^{2}}$. For these
values of $u$ there is a tradeoff: higher $u$ gives a larger transition
metal-sulfur distance but a smaller sulfur-sulfur distance. These
competing effects lead to the clear non-monotonic behavior of $u$.
For less than $0.25$ added electrons per f.u. the effect on the sulfur-sulfur
distance is more important, and $u$ decreases. However, for larger
numbers of added electrons, once the $e_{g}$ states begin filling,
the transition metal-sulfur distance becomes more important and $u$ increases.

\section{Tight-Binding Model}

\label{sec:Tight-Binding-Model}

While DFT is able to determine the ground state energy of a specific
magnetic configuration, testing all possible types of magnetic order
to find the lowest energy state is infeasible. Instead, to screen
the possible magnetic wave-vectors, we compute the non-interacting
magnetic susceptibility via the Lindhard function. In a weakly interacting
system, which should describe doped \chem{FeS_2}, this quantity
provides a good indicator of the different instabilities of the system.
While more sophisticated calculations that account for electronic
interactions are possible, for the scope of this work it suffices to
consider the non-interacting susceptibility.

\begin{figure}
\includegraphics[width=0.5\textwidth]{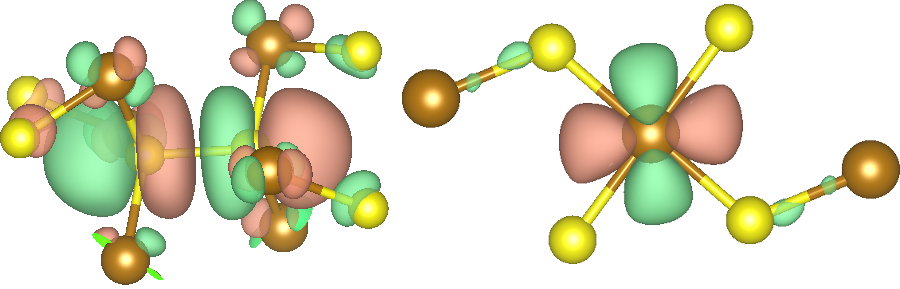} 
\caption{Illustraton of the sulfur-dimer centered Wannier function (left) and transition metal centered Wannier function (right). The sulfur-dimer function has $p$ anti-bonding character, which is the character of the wide band at the bottom of the conduction band shown in Fig. \ref{fig:schematicDOS}(a).}
\label{fig:SwannierFunc} 
\end{figure}

To efficiently compute the Lindhard function, we first construct a
tight-binding model from the Wannier functions, which are obtained
from a unitary transformation of the Bloch wavefunctions into a new
basis. The resulting functions are maximally spatially localized and
entirely real \cite{wannierTheory}. In practice, we calculate an approximate
unitary transformation that minimizes the real space spread of the
wavefunctions. This additionally gives a maximum ratio of the real
and imaginary parts of the wavefunction of less than $10^{-5}$. Wannier
models are regularly used to interpolate band structures  \cite{ramasubramaniam_large_2012},
to map Fermi surfaces \cite{shekhar_extremely_2015}, and to calculate
Fermi surface integrals \cite{wanniersFSintegrals}. Figure \ref{fig:SwannierFunc}
shows an example of a Wannier function centered on a sulfur dimer
with some hybridization to the six transition metal atoms neighboring
the sulfur dimer. The other Wannier functions have well-localized
$e_{g}$ character on the transition metal atoms, as also shown in Fig. \ref{fig:SwannierFunc}.
This Wannier function further emphasizes the covalency between the S atoms in dimers, 
since the function is centered at the bond center and not on an individual 
S ion.
These dimer orbitals are important to the overall band structure,
as discussed above. The entire conduction manifold that consists of the transition metal $e_{g}$ and 
sulfur anti-bonding states is used for our Wannier calculations, generating twelve Wannier functions per cell, 
with no need for disentanglement.

These functions allow us to efficiently derive a tight-binding model
of the form: 
\begin{equation}
H=\sum_{\vec{R},st}t_{st}^{\vec{R}}\left(\hat{c}_{\vec{R},s}^{\dagger}\hat{c}_{\vec{0},t}+\text{c.c.}\right).
\end{equation}
from the Wannier basis, where $\hat{c}^{\dagger},\hat{c}$ are creation
and annihilation operators, $\vec{R}$ is the vector connecting the
unit cells of two orbitals, and $s,t$ are orbital indices within
a cell (spin indices are omitted for simplicity). The hopping terms
$t_{st}^{\vec{R}}=\braket{\vec{0}s|\hat{H}|\vec{R}t}$ are directly
computed as the matrix element between the $s^{th}$ Wannier function
in the home cell and the $t^{th}$ Wannier function in the cell at $\vec{R}$.
Because we are not interested in finding a minimal model, twenty distinct
hopping vectors are kept corresponding to approximately 700 separate
terms. This large number of terms allows us to obtain almost exact
agreement with the DFT band structure for all bands (see the Appendix
for all DFT and tight binding bandstructures). With this model we
can very efficiently compute energies at arbitrary k-points.

From the tight-binding model we calculate the magnetic susceptibility
in the first Brillouin zone by computing the Lindhardt function. The
general non-interacting magnetic susceptibility is given by \cite{chi0}

\begin{align}
\chi_{st}^{pq}({\mathbf q},\omega)\!=\!
-\frac{1}{N}\!\!\sum_{{\mathbf k},\mu\nu}\!\! 
& \frac{
a_{\mu}^{s}({\mathbf k})a_{\mu}^{p*}({\mathbf k})a_{\nu}^{q}({\mathbf k}+{\mathbf q})a_{\nu}^{t*}({\mathbf k}+{\mathbf q})}
{\omega+E_{\nu}({\mathbf k}+{\mathbf q})-E_{\mu}({\mathbf k})+i0^{+}}\\
 & \times[f(E_{\nu}({\mathbf k}+{\mathbf q}))-f(E_{\mu}({\mathbf k}))],\nonumber 
\end{align}
where $a^s_\mu({\mathbf k})$ are the matrix elements corresponding to the change from orbital basis (latin indices) to band basis (greek indices), $E_\mu({\mathbf k})$ is the energy of band $\mu$ at momentum ${\mathbf k}$, and $N$ is the number of sites. The static susceptibility is \cite{chi0}
\begin{align}
\chi^{0}({\mathbf q})=\frac{1}{2}\sum_{sp}\chi_{ss}^{pp}({\mathbf q},0).
\end{align}

\begin{figure}
\includegraphics[width=0.45\textwidth]{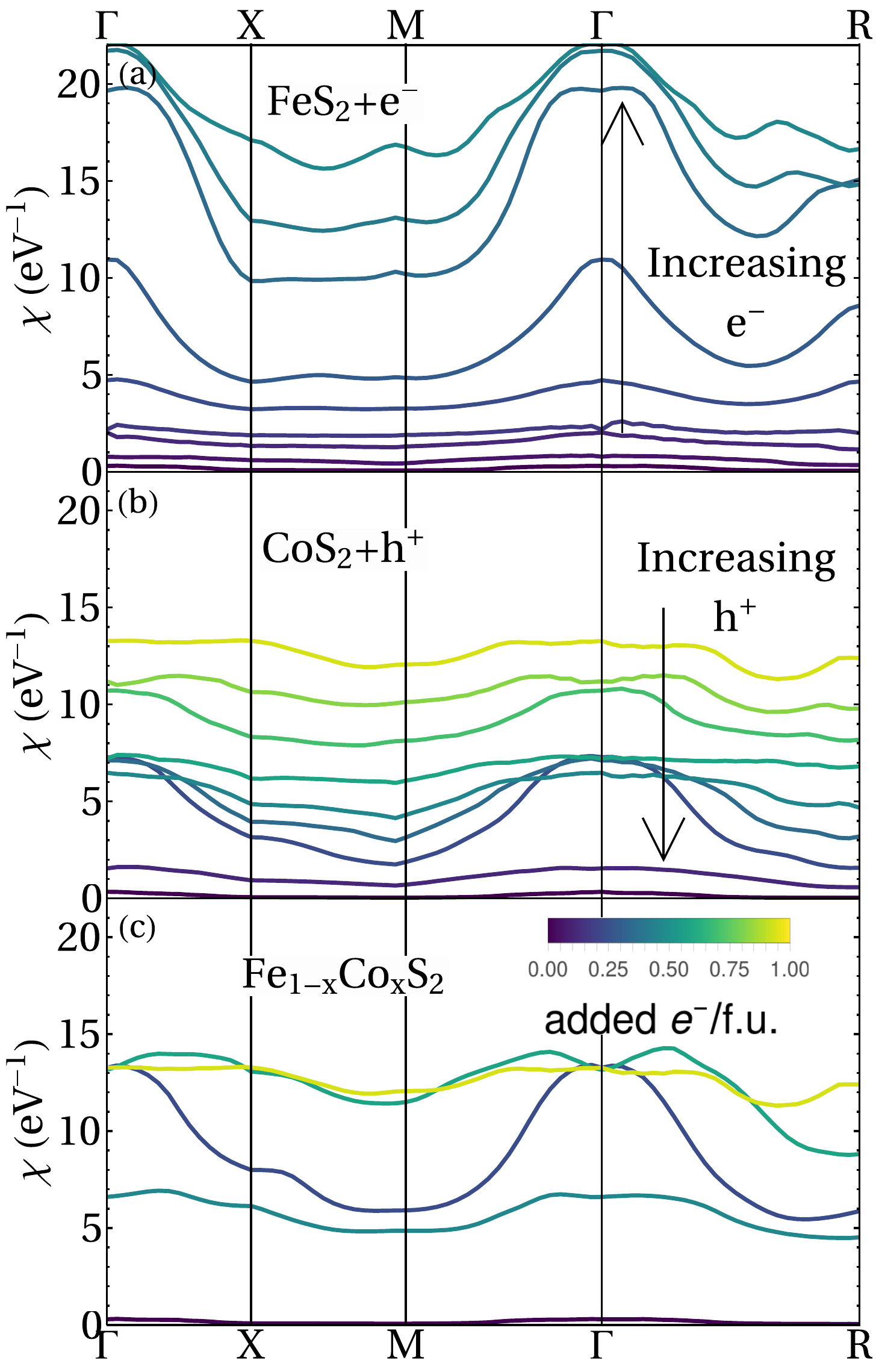} \caption{Non-interacting magnetic susceptibilities in momentum space $\chi(q)$
for (a)gated \chem{FeS_2}, (b) gated \chem{CoS_2}, and (c) doped
\chem{Fe_{1-x}Co_xS_2}. Note how in (c) the overall magnitude of
the curves is not monotonic with doping.}
\label{fig:chis} 
\end{figure}

Figure \ref{fig:chis} shows the susceptibilities for EG \chem{FeS_2}
and \chem{CoS_2}, as well as for CD \chem{Fe_{1-x}Co_xS_2}. In agreement
with the spin-polarized DFT calculations, we observe a sharp increase
in the magnetic susceptibility at the $\Gamma$ point (i.e. $\mathbf{q}=0$)
starting at $0.25$ added electrons per f.u. in both gated compounds,
consistent with a tendency towards ferromagnetism. Note that $\chi_{0}(\Gamma)$
is proportional to the density of states at the Fermi level; thus,
the non-monotonic behavior of the magnitude of $\chi(q)$ as function
of doping in the case of CD \chem{Fe_{1-x}Co_xS_2} is consistent
with the non-monotonic dependence of the DOS shown in Fig. \ref{fig:gamma}.
The key point of this calculation is to show that, when ferromagnetism
emerges upon adding 0.25 electrons per formula unit, the non-interacting susceptibility displays no competing 
peaks at other wave-vectors. This makes it less likely that competing
magnetic states are realized in this compound and, moreover, lends
support to the proposal that the ferromagnetism is of Stoner-type.

\section{Conclusions}

\label{sec:Conclusions-and-Summary}

We performed first principles calculations for both chemically doped
\chem{Fe_{1-x}Co_xS_2} and electrostatically gated \chem{FeS_2}
and \chem{CoS_2} to elucidate how these different ways of changing the carreir
concentration affect the magnetic and electronic properties of these
pyrite compounds. We found that electrostatic gating requires a larger concentration of added electrons to induce ferromagnetism as compared to chemical doping. We attribute this behavior to the Stoner nature 
of the ferromagnetic instability, combined with the fundamentally
different ways in which the band structure changes upon gating versus
doping. Specifically, while \chem{Co} $e_{g}$ bands with large
DOS form at the bottom of the conduction band when FeS$_{2}$ is doped
with Co, these bands are not present in electrostatically gated FeS$_{2}$.
Instead, in the latter case, a low DOS wide sulfur band must first
be occupied before the $e_{g}$ Fe band becomes filled, thus delaying
the onset of ferromagnetism.

Our structural relaxation calculations revealed significant changes
in several relevant crystalline parameters upon adding electrons via
gating. This result demonstrates that electrostatic gating has a much
richer impact beyond a rigid-band shift, altering both the crystal
structure and the electronic structure. Finally, our tight-binding
parametrization allowed us to compute the non-interacting magnetic
susceptibility, which revealed a sharp peak at the $\Gamma$ point consistent
with a leading ferromagnetic Stoner-like instability.

Our investigation shows that, even without considering the impact
of disorder introduced by dopants, electrostatic gating and chemical
doping can affect the electronic properties of a compound in rather
different ways, resulting in distinct macroscopic properties. This
also suggests that a combination of electrostatic gating and chemical
doping may provide an interesting and efficient way to probe and tune
electronic ground states. 
We note that
the current capabilities of ionic liquid or gel gating of adding $10^{14}\text{cm}^{-2}$ would correspond to adding $\approx 0.3$ electrons per formula unit in the case of \chem{FeS_2} (assuming a penetration depth of one unit cell). Thus, our results suggest that electrolyte gating is a viable means to induce ferromagnetism in \chem{FeS_2}
purely electrostatically.

\acknowledgements This work was supported by the National Science Foundation through the UMN MRSEC under DMR-1420013. The authors acknowledge the Minnesota Supercomputing Institute (MSI) at the University of Minnesota for providing resources that contributed to the research results reported within this paper.

\bibliographystyle{apsrev4-1}

\begin{thebibliography}{78}%
\makeatletter
\providecommand \@ifxundefined [1]{%
 \@ifx{#1\undefined}
}%
\providecommand \@ifnum [1]{%
 \ifnum #1\expandafter \@firstoftwo
 \else \expandafter \@secondoftwo
 \fi
}%
\providecommand \@ifx [1]{%
 \ifx #1\expandafter \@firstoftwo
 \else \expandafter \@secondoftwo
 \fi
}%
\providecommand \natexlab [1]{#1}%
\providecommand \enquote  [1]{``#1''}%
\providecommand \bibnamefont  [1]{#1}%
\providecommand \bibfnamefont [1]{#1}%
\providecommand \citenamefont [1]{#1}%
\providecommand \href@noop [0]{\@secondoftwo}%
\providecommand \href [0]{\begingroup \@sanitize@url \@href}%
\providecommand \@href[1]{\@@startlink{#1}\@@href}%
\providecommand \@@href[1]{\endgroup#1\@@endlink}%
\providecommand \@sanitize@url [0]{\catcode `\\12\catcode `\$12\catcode
  `\&12\catcode `\#12\catcode `\^12\catcode `\_12\catcode `\%12\relax}%
\providecommand \@@startlink[1]{}%
\providecommand \@@endlink[0]{}%
\providecommand \url  [0]{\begingroup\@sanitize@url \@url }%
\providecommand \@url [1]{\endgroup\@href {#1}{\urlprefix }}%
\providecommand \urlprefix  [0]{URL }%
\providecommand \Eprint [0]{\href }%
\providecommand \doibase [0]{http://dx.doi.org/}%
\providecommand \selectlanguage [0]{\@gobble}%
\providecommand \bibinfo  [0]{\@secondoftwo}%
\providecommand \bibfield  [0]{\@secondoftwo}%
\providecommand \translation [1]{[#1]}%
\providecommand \BibitemOpen [0]{}%
\providecommand \bibitemStop [0]{}%
\providecommand \bibitemNoStop [0]{.\EOS\space}%
\providecommand \EOS [0]{\spacefactor3000\relax}%
\providecommand \BibitemShut  [1]{\csname bibitem#1\endcsname}%
\let\auto@bib@innerbib\@empty
\bibitem [{\citenamefont {Ogawa}(1976)}]{Ogawa1976}%
  \BibitemOpen
  \bibfield  {author} {\bibinfo {author} {\bibfnamefont {S.}~\bibnamefont
  {Ogawa}},\ }\href {\doibase 10.1143/JPSJ.41.462} {\bibfield  {journal}
  {\bibinfo  {journal} {Journal of the Physical Society of Japan}\ }\textbf
  {\bibinfo {volume} {41}},\ \bibinfo {pages} {462} (\bibinfo {year}
  {1976})}\BibitemShut {NoStop}%
\bibitem [{\citenamefont {Chattopadhyay}\ \emph {et~al.}(1991)\citenamefont
  {Chattopadhyay}, \citenamefont {Burlet},\ and\ \citenamefont
  {Brown}}]{Chattopadhyay_1991}%
  \BibitemOpen
  \bibfield  {author} {\bibinfo {author} {\bibfnamefont {T.}~\bibnamefont
  {Chattopadhyay}}, \bibinfo {author} {\bibfnamefont {P.}~\bibnamefont
  {Burlet}}, \ and\ \bibinfo {author} {\bibfnamefont {P.~J.}\ \bibnamefont
  {Brown}},\ }\href {\doibase 10.1088/0953-8984/3/29/009} {\bibfield  {journal}
  {\bibinfo  {journal} {Journal of Physics: Condensed Matter}\ }\textbf
  {\bibinfo {volume} {3}},\ \bibinfo {pages} {5555} (\bibinfo {year}
  {1991})}\BibitemShut {NoStop}%
\bibitem [{\citenamefont {Lin}\ and\ \citenamefont
  {Hacker}(1968)}]{LIN1968687}%
  \BibitemOpen
  \bibfield  {author} {\bibinfo {author} {\bibfnamefont {M.}~\bibnamefont
  {Lin}}\ and\ \bibinfo {author} {\bibfnamefont {H.}~\bibnamefont {Hacker}},\
  }\href {\doibase https://doi.org/10.1016/0038-1098(68)90564-4} {\bibfield
  {journal} {\bibinfo  {journal} {Solid State Communications}\ }\textbf
  {\bibinfo {volume} {6}},\ \bibinfo {pages} {687 } (\bibinfo {year}
  {1968})}\BibitemShut {NoStop}%
\bibitem [{\citenamefont {Kimber}\ and\ \citenamefont
  {Chatterji}(2015)}]{Kimber_2015}%
  \BibitemOpen
  \bibfield  {author} {\bibinfo {author} {\bibfnamefont {S.~A.~J.}\
  \bibnamefont {Kimber}}\ and\ \bibinfo {author} {\bibfnamefont
  {T.}~\bibnamefont {Chatterji}},\ }\href {\doibase
  10.1088/0953-8984/27/22/226003} {\bibfield  {journal} {\bibinfo  {journal}
  {Journal of Physics: Condensed Matter}\ }\textbf {\bibinfo {volume} {27}},\
  \bibinfo {pages} {226003} (\bibinfo {year} {2015})}\BibitemShut {NoStop}%
\bibitem [{\citenamefont {Zhao}\ \emph {et~al.}(1993)\citenamefont {Zhao},
  \citenamefont {Callaway},\ and\ \citenamefont
  {Hayashibara}}]{PhysRevB.48.15781}%
  \BibitemOpen
  \bibfield  {author} {\bibinfo {author} {\bibfnamefont {G.~L.}\ \bibnamefont
  {Zhao}}, \bibinfo {author} {\bibfnamefont {J.}~\bibnamefont {Callaway}}, \
  and\ \bibinfo {author} {\bibfnamefont {M.}~\bibnamefont {Hayashibara}},\
  }\href {\doibase 10.1103/PhysRevB.48.15781} {\bibfield  {journal} {\bibinfo
  {journal} {Phys. Rev. B}\ }\textbf {\bibinfo {volume} {48}},\ \bibinfo
  {pages} {15781} (\bibinfo {year} {1993})}\BibitemShut {NoStop}%
\bibitem [{\citenamefont {Leighton}\ \emph {et~al.}(2007)\citenamefont
  {Leighton}, \citenamefont {Manno}, \citenamefont {Cady}, \citenamefont
  {Freeland}, \citenamefont {Wang}, \citenamefont {Umemoto}, \citenamefont
  {Wentzcovitch}, \citenamefont {Chen}, \citenamefont {Chien}, \citenamefont
  {Kuhns}, \citenamefont {Hoch}, \citenamefont {Reyes}, \citenamefont
  {Moulton}, \citenamefont {Dahlberg}, \citenamefont {Checkelsky},\ and\
  \citenamefont {Eckert}}]{leighton_composition_2007}%
  \BibitemOpen
  \bibfield  {author} {\bibinfo {author} {\bibfnamefont {C.}~\bibnamefont
  {Leighton}}, \bibinfo {author} {\bibfnamefont {M.}~\bibnamefont {Manno}},
  \bibinfo {author} {\bibfnamefont {A.}~\bibnamefont {Cady}}, \bibinfo {author}
  {\bibfnamefont {J.~W.}\ \bibnamefont {Freeland}}, \bibinfo {author}
  {\bibfnamefont {L.}~\bibnamefont {Wang}}, \bibinfo {author} {\bibfnamefont
  {K.}~\bibnamefont {Umemoto}}, \bibinfo {author} {\bibfnamefont {R.~M.}\
  \bibnamefont {Wentzcovitch}}, \bibinfo {author} {\bibfnamefont {T.~Y.}\
  \bibnamefont {Chen}}, \bibinfo {author} {\bibfnamefont {C.~L.}\ \bibnamefont
  {Chien}}, \bibinfo {author} {\bibfnamefont {P.~L.}\ \bibnamefont {Kuhns}},
  \bibinfo {author} {\bibfnamefont {M.~J.~R.}\ \bibnamefont {Hoch}}, \bibinfo
  {author} {\bibfnamefont {A.~P.}\ \bibnamefont {Reyes}}, \bibinfo {author}
  {\bibfnamefont {W.~G.}\ \bibnamefont {Moulton}}, \bibinfo {author}
  {\bibfnamefont {E.~D.}\ \bibnamefont {Dahlberg}}, \bibinfo {author}
  {\bibfnamefont {J.}~\bibnamefont {Checkelsky}}, \ and\ \bibinfo {author}
  {\bibfnamefont {J.}~\bibnamefont {Eckert}},\ }\href {\doibase
  10.1088/0953-8984/19/31/315219} {\bibfield  {journal} {\bibinfo  {journal}
  {Journal of Physics: Condensed Matter}\ }\textbf {\bibinfo {volume} {19}},\
  \bibinfo {pages} {315219} (\bibinfo {year} {2007})}\BibitemShut {NoStop}%
\bibitem [{\citenamefont {Matsuura}\ \emph {et~al.}(2003)\citenamefont
  {Matsuura}, \citenamefont {Endoh}, \citenamefont {Hiraka}, \citenamefont
  {Yamada}, \citenamefont {Mishchenko}, \citenamefont {Nagaosa},\ and\
  \citenamefont {Solovyev}}]{PhysRevB.68.094409}%
  \BibitemOpen
  \bibfield  {author} {\bibinfo {author} {\bibfnamefont {M.}~\bibnamefont
  {Matsuura}}, \bibinfo {author} {\bibfnamefont {Y.}~\bibnamefont {Endoh}},
  \bibinfo {author} {\bibfnamefont {H.}~\bibnamefont {Hiraka}}, \bibinfo
  {author} {\bibfnamefont {K.}~\bibnamefont {Yamada}}, \bibinfo {author}
  {\bibfnamefont {A.~S.}\ \bibnamefont {Mishchenko}}, \bibinfo {author}
  {\bibfnamefont {N.}~\bibnamefont {Nagaosa}}, \ and\ \bibinfo {author}
  {\bibfnamefont {I.~V.}\ \bibnamefont {Solovyev}},\ }\href {\doibase
  10.1103/PhysRevB.68.094409} {\bibfield  {journal} {\bibinfo  {journal} {Phys.
  Rev. B}\ }\textbf {\bibinfo {volume} {68}},\ \bibinfo {pages} {094409}
  (\bibinfo {year} {2003})}\BibitemShut {NoStop}%
\bibitem [{\citenamefont {Schuster}\ \emph {et~al.}(2012)\citenamefont
  {Schuster}, \citenamefont {Gatti},\ and\ \citenamefont
  {Rubio}}]{Schuster2012}%
  \BibitemOpen
  \bibfield  {author} {\bibinfo {author} {\bibfnamefont {C.}~\bibnamefont
  {Schuster}}, \bibinfo {author} {\bibfnamefont {M.}~\bibnamefont {Gatti}}, \
  and\ \bibinfo {author} {\bibfnamefont {A.}~\bibnamefont {Rubio}},\ }\href
  {\doibase 10.1140/epjb/e2012-30384-7} {\bibfield  {journal} {\bibinfo
  {journal} {The European Physical Journal B}\ }\textbf {\bibinfo {volume}
  {85}},\ \bibinfo {pages} {325} (\bibinfo {year} {2012})}\BibitemShut
  {NoStop}%
\bibitem [{\citenamefont {Yano}\ \emph {et~al.}(2016)\citenamefont {Yano},
  \citenamefont {Louca}, \citenamefont {Yang}, \citenamefont {Chatterjee},
  \citenamefont {Bugaris}, \citenamefont {Chung}, \citenamefont {Peng},
  \citenamefont {Grayson},\ and\ \citenamefont {Kanatzidis}}]{Yano2016}%
  \BibitemOpen
  \bibfield  {author} {\bibinfo {author} {\bibfnamefont {S.}~\bibnamefont
  {Yano}}, \bibinfo {author} {\bibfnamefont {D.}~\bibnamefont {Louca}},
  \bibinfo {author} {\bibfnamefont {J.}~\bibnamefont {Yang}}, \bibinfo {author}
  {\bibfnamefont {U.}~\bibnamefont {Chatterjee}}, \bibinfo {author}
  {\bibfnamefont {D.~E.}\ \bibnamefont {Bugaris}}, \bibinfo {author}
  {\bibfnamefont {D.~Y.}\ \bibnamefont {Chung}}, \bibinfo {author}
  {\bibfnamefont {L.}~\bibnamefont {Peng}}, \bibinfo {author} {\bibfnamefont
  {M.}~\bibnamefont {Grayson}}, \ and\ \bibinfo {author} {\bibfnamefont
  {M.~G.}\ \bibnamefont {Kanatzidis}},\ }\href {\doibase
  10.1103/PhysRevB.93.024409} {\bibfield  {journal} {\bibinfo  {journal} {Phys.
  Rev. B}\ }\textbf {\bibinfo {volume} {93}},\ \bibinfo {pages} {024409}
  (\bibinfo {year} {2016})}\BibitemShut {NoStop}%
\bibitem [{\citenamefont {Bither}\ \emph {et~al.}(1966)\citenamefont {Bither},
  \citenamefont {Prewitt}, \citenamefont {Gillson}, \citenamefont {Bierstedt},
  \citenamefont {Flippen},\ and\ \citenamefont {Young}}]{BITHER1966533}%
  \BibitemOpen
  \bibfield  {author} {\bibinfo {author} {\bibfnamefont {T.}~\bibnamefont
  {Bither}}, \bibinfo {author} {\bibfnamefont {C.}~\bibnamefont {Prewitt}},
  \bibinfo {author} {\bibfnamefont {J.}~\bibnamefont {Gillson}}, \bibinfo
  {author} {\bibfnamefont {P.}~\bibnamefont {Bierstedt}}, \bibinfo {author}
  {\bibfnamefont {R.}~\bibnamefont {Flippen}}, \ and\ \bibinfo {author}
  {\bibfnamefont {H.}~\bibnamefont {Young}},\ }\href {\doibase
  https://doi.org/10.1016/0038-1098(66)90419-4} {\bibfield  {journal} {\bibinfo
   {journal} {Solid State Communications}\ }\textbf {\bibinfo {volume} {4}},\
  \bibinfo {pages} {533 } (\bibinfo {year} {1966})}\BibitemShut {NoStop}%
\bibitem [{\citenamefont {Ueda}\ \emph {et~al.}(2002)\citenamefont {Ueda},
  \citenamefont {Nohara}, \citenamefont {Kitazawa}, \citenamefont {Takagi},
  \citenamefont {Fujimori}, \citenamefont {Mizokawa},\ and\ \citenamefont
  {Yagi}}]{PhysRevB.65.155104}%
  \BibitemOpen
  \bibfield  {author} {\bibinfo {author} {\bibfnamefont {H.}~\bibnamefont
  {Ueda}}, \bibinfo {author} {\bibfnamefont {M.}~\bibnamefont {Nohara}},
  \bibinfo {author} {\bibfnamefont {K.}~\bibnamefont {Kitazawa}}, \bibinfo
  {author} {\bibfnamefont {H.}~\bibnamefont {Takagi}}, \bibinfo {author}
  {\bibfnamefont {A.}~\bibnamefont {Fujimori}}, \bibinfo {author}
  {\bibfnamefont {T.}~\bibnamefont {Mizokawa}}, \ and\ \bibinfo {author}
  {\bibfnamefont {T.}~\bibnamefont {Yagi}},\ }\href {\doibase
  10.1103/PhysRevB.65.155104} {\bibfield  {journal} {\bibinfo  {journal} {Phys.
  Rev. B}\ }\textbf {\bibinfo {volume} {65}},\ \bibinfo {pages} {155104}
  (\bibinfo {year} {2002})}\BibitemShut {NoStop}%
\bibitem [{\citenamefont {Bullett}(1982)}]{Bullett_1982}%
  \BibitemOpen
  \bibfield  {author} {\bibinfo {author} {\bibfnamefont {D.~W.}\ \bibnamefont
  {Bullett}},\ }\href {\doibase 10.1088/0022-3719/15/30/010} {\bibfield
  {journal} {\bibinfo  {journal} {Journal of Physics C: Solid State Physics}\
  }\textbf {\bibinfo {volume} {15}},\ \bibinfo {pages} {6163} (\bibinfo {year}
  {1982})}\BibitemShut {NoStop}%
\bibitem [{\citenamefont {Goldman}(2014)}]{goldman_electrostatic_2014}%
  \BibitemOpen
  \bibfield  {author} {\bibinfo {author} {\bibfnamefont {A.}~\bibnamefont
  {Goldman}},\ }\href {\doibase 10.1146/annurev-matsci-070813-113407}
  {\bibfield  {journal} {\bibinfo  {journal} {Annu. Rev. Mater. Res.}\ }\textbf
  {\bibinfo {volume} {44}},\ \bibinfo {pages} {45} (\bibinfo {year}
  {2014})}\BibitemShut {NoStop}%
\bibitem [{\citenamefont {Burton}\ and\ \citenamefont
  {Tsymbal}(2011)}]{Burton2011}%
  \BibitemOpen
  \bibfield  {author} {\bibinfo {author} {\bibfnamefont {J.~D.}\ \bibnamefont
  {Burton}}\ and\ \bibinfo {author} {\bibfnamefont {E.~Y.}\ \bibnamefont
  {Tsymbal}},\ }\href {\doibase 10.1103/PhysRevLett.107.166601} {\bibfield
  {journal} {\bibinfo  {journal} {Physical Review Letters}\ }\textbf {\bibinfo
  {volume} {107}},\ \bibinfo {pages} {166601} (\bibinfo {year}
  {2011})}\BibitemShut {NoStop}%
\bibitem [{\citenamefont {Liu}\ \emph {et~al.}(2018)\citenamefont {Liu},
  \citenamefont {Tsymbal},\ and\ \citenamefont {Rabe}}]{Liu2018}%
  \BibitemOpen
  \bibfield  {author} {\bibinfo {author} {\bibfnamefont {X.}~\bibnamefont
  {Liu}}, \bibinfo {author} {\bibfnamefont {E.~Y.}\ \bibnamefont {Tsymbal}}, \
  and\ \bibinfo {author} {\bibfnamefont {K.~M.}\ \bibnamefont {Rabe}},\ }\href
  {\doibase 10.1103/PhysRevB.97.094107} {\bibfield  {journal} {\bibinfo
  {journal} {Physical Review B}\ }\textbf {\bibinfo {volume} {97}},\ \bibinfo
  {pages} {094107} (\bibinfo {year} {2018})}\BibitemShut {NoStop}%
\bibitem [{\citenamefont {Zhang}\ \emph {et~al.}(2019)\citenamefont {Zhang},
  \citenamefont {Zhang}, \citenamefont {Zhou}, \citenamefont {Tanaka},
  \citenamefont {Fong},\ and\ \citenamefont {Ramanathan}}]{Zhang1523686}%
  \BibitemOpen
  \bibfield  {author} {\bibinfo {author} {\bibfnamefont {H.-T.}\ \bibnamefont
  {Zhang}}, \bibinfo {author} {\bibfnamefont {Z.}~\bibnamefont {Zhang}},
  \bibinfo {author} {\bibfnamefont {H.}~\bibnamefont {Zhou}}, \bibinfo {author}
  {\bibfnamefont {H.}~\bibnamefont {Tanaka}}, \bibinfo {author} {\bibfnamefont
  {D.~D.}\ \bibnamefont {Fong}}, \ and\ \bibinfo {author} {\bibfnamefont
  {S.}~\bibnamefont {Ramanathan}},\ }\href {\doibase
  10.1080/23746149.2018.1523686} {\bibfield  {journal} {\bibinfo  {journal}
  {Advances in Physics: X}\ }\textbf {\bibinfo {volume} {4}},\ \bibinfo {pages}
  {1523686} (\bibinfo {year} {2019})}\BibitemShut {NoStop}%
\bibitem [{\citenamefont {Song}\ \emph {et~al.}(2016)\citenamefont {Song},
  \citenamefont {Cui}, \citenamefont {Peng}, \citenamefont {Mao},\ and\
  \citenamefont {Pan}}]{Song_2016}%
  \BibitemOpen
  \bibfield  {author} {\bibinfo {author} {\bibfnamefont {C.}~\bibnamefont
  {Song}}, \bibinfo {author} {\bibfnamefont {B.}~\bibnamefont {Cui}}, \bibinfo
  {author} {\bibfnamefont {J.}~\bibnamefont {Peng}}, \bibinfo {author}
  {\bibfnamefont {H.}~\bibnamefont {Mao}}, \ and\ \bibinfo {author}
  {\bibfnamefont {F.}~\bibnamefont {Pan}},\ }\href {\doibase
  10.1088/1674-1056/25/6/067502} {\bibfield  {journal} {\bibinfo  {journal}
  {Chinese Physics B}\ }\textbf {\bibinfo {volume} {25}},\ \bibinfo {pages}
  {067502} (\bibinfo {year} {2016})}\BibitemShut {NoStop}%
\bibitem [{\citenamefont {Scherwitzl}\ \emph {et~al.}(2010)\citenamefont
  {Scherwitzl}, \citenamefont {Zubko}, \citenamefont {Lezama}, \citenamefont
  {Ono}, \citenamefont {Morpurgo}, \citenamefont {Catalan},\ and\ \citenamefont
  {Triscone}}]{Scherwitzl2010}%
  \BibitemOpen
  \bibfield  {author} {\bibinfo {author} {\bibfnamefont {R.}~\bibnamefont
  {Scherwitzl}}, \bibinfo {author} {\bibfnamefont {P.}~\bibnamefont {Zubko}},
  \bibinfo {author} {\bibfnamefont {I.~G.}\ \bibnamefont {Lezama}}, \bibinfo
  {author} {\bibfnamefont {S.}~\bibnamefont {Ono}}, \bibinfo {author}
  {\bibfnamefont {A.~F.}\ \bibnamefont {Morpurgo}}, \bibinfo {author}
  {\bibfnamefont {G.}~\bibnamefont {Catalan}}, \ and\ \bibinfo {author}
  {\bibfnamefont {J.-M.}\ \bibnamefont {Triscone}},\ }\href {\doibase
  10.1002/adma.201003241} {\bibfield  {journal} {\bibinfo  {journal} {Advanced
  Materials}\ }\textbf {\bibinfo {volume} {22}},\ \bibinfo {pages} {5517}
  (\bibinfo {year} {2010})}\BibitemShut {NoStop}%
\bibitem [{\citenamefont {Ueno}\ \emph {et~al.}(2008)\citenamefont {Ueno},
  \citenamefont {Nakamura}, \citenamefont {Shimotani}, \citenamefont {Ohtomo},
  \citenamefont {Kimura}, \citenamefont {Nojima}, \citenamefont {Aoki},
  \citenamefont {Iwasa},\ and\ \citenamefont {Kawasaki}}]{Ueno2008}%
  \BibitemOpen
  \bibfield  {author} {\bibinfo {author} {\bibfnamefont {K.}~\bibnamefont
  {Ueno}}, \bibinfo {author} {\bibfnamefont {S.}~\bibnamefont {Nakamura}},
  \bibinfo {author} {\bibfnamefont {H.}~\bibnamefont {Shimotani}}, \bibinfo
  {author} {\bibfnamefont {A.}~\bibnamefont {Ohtomo}}, \bibinfo {author}
  {\bibfnamefont {N.}~\bibnamefont {Kimura}}, \bibinfo {author} {\bibfnamefont
  {T.}~\bibnamefont {Nojima}}, \bibinfo {author} {\bibfnamefont
  {H.}~\bibnamefont {Aoki}}, \bibinfo {author} {\bibfnamefont {Y.}~\bibnamefont
  {Iwasa}}, \ and\ \bibinfo {author} {\bibfnamefont {M.}~\bibnamefont
  {Kawasaki}},\ }\href {\doibase 10.1038/nmat2298} {\bibfield  {journal}
  {\bibinfo  {journal} {Nature Materials}\ }\textbf {\bibinfo {volume} {7}},\
  \bibinfo {pages} {855} (\bibinfo {year} {2008})}\BibitemShut {NoStop}%
\bibitem [{\citenamefont {Jeong}\ \emph {et~al.}(2013)\citenamefont {Jeong},
  \citenamefont {Aetukuri}, \citenamefont {Graf}, \citenamefont {Schladt},
  \citenamefont {Samant},\ and\ \citenamefont {Parkin}}]{Jeong2013}%
  \BibitemOpen
  \bibfield  {author} {\bibinfo {author} {\bibfnamefont {J.}~\bibnamefont
  {Jeong}}, \bibinfo {author} {\bibfnamefont {N.}~\bibnamefont {Aetukuri}},
  \bibinfo {author} {\bibfnamefont {T.}~\bibnamefont {Graf}}, \bibinfo {author}
  {\bibfnamefont {T.~D.}\ \bibnamefont {Schladt}}, \bibinfo {author}
  {\bibfnamefont {M.~G.}\ \bibnamefont {Samant}}, \ and\ \bibinfo {author}
  {\bibfnamefont {S.~S.~P.}\ \bibnamefont {Parkin}},\ }\href {\doibase
  10.1126/science.1230512} {\bibfield  {journal} {\bibinfo  {journal}
  {Science}\ }\textbf {\bibinfo {volume} {339}},\ \bibinfo {pages} {1402}
  (\bibinfo {year} {2013})}\BibitemShut {NoStop}%
\bibitem [{\citenamefont {Bisri}\ \emph {et~al.}(2017)\citenamefont {Bisri},
  \citenamefont {Shimizu}, \citenamefont {Nakano},\ and\ \citenamefont
  {Iwasa}}]{Bisri2017}%
  \BibitemOpen
  \bibfield  {author} {\bibinfo {author} {\bibfnamefont {S.~Z.}\ \bibnamefont
  {Bisri}}, \bibinfo {author} {\bibfnamefont {S.}~\bibnamefont {Shimizu}},
  \bibinfo {author} {\bibfnamefont {M.}~\bibnamefont {Nakano}}, \ and\ \bibinfo
  {author} {\bibfnamefont {Y.}~\bibnamefont {Iwasa}},\ }\href {\doibase
  10.1002/adma.201607054} {\bibfield  {journal} {\bibinfo  {journal} {Advanced
  Materials}\ }\textbf {\bibinfo {volume} {29}},\ \bibinfo {pages} {1607054}
  (\bibinfo {year} {2017})}\BibitemShut {NoStop}%
\bibitem [{\citenamefont {Yuan}\ \emph {et~al.}(2009)\citenamefont {Yuan},
  \citenamefont {Shimotani}, \citenamefont {Tsukazaki}, \citenamefont {Ohtomo},
  \citenamefont {Kawasaki},\ and\ \citenamefont
  {Iwasa}}]{10.1002/adfm.200801633}%
  \BibitemOpen
  \bibfield  {author} {\bibinfo {author} {\bibfnamefont {H.}~\bibnamefont
  {Yuan}}, \bibinfo {author} {\bibfnamefont {H.}~\bibnamefont {Shimotani}},
  \bibinfo {author} {\bibfnamefont {A.}~\bibnamefont {Tsukazaki}}, \bibinfo
  {author} {\bibfnamefont {A.}~\bibnamefont {Ohtomo}}, \bibinfo {author}
  {\bibfnamefont {M.}~\bibnamefont {Kawasaki}}, \ and\ \bibinfo {author}
  {\bibfnamefont {Y.}~\bibnamefont {Iwasa}},\ }\href {\doibase
  10.1002/adfm.200801633} {\bibfield  {journal} {\bibinfo  {journal} {Advanced
  Functional Materials}\ }\textbf {\bibinfo {volume} {19}},\ \bibinfo {pages}
  {1046} (\bibinfo {year} {2009})}\BibitemShut {NoStop}%
\bibitem [{\citenamefont {Jeong}\ \emph {et~al.}(2015)\citenamefont {Jeong},
  \citenamefont {Aetukuri}, \citenamefont {Passarello}, \citenamefont
  {Conradson}, \citenamefont {Samant},\ and\ \citenamefont
  {Parkin}}]{Jeong1013}%
  \BibitemOpen
  \bibfield  {author} {\bibinfo {author} {\bibfnamefont {J.}~\bibnamefont
  {Jeong}}, \bibinfo {author} {\bibfnamefont {N.~B.}\ \bibnamefont {Aetukuri}},
  \bibinfo {author} {\bibfnamefont {D.}~\bibnamefont {Passarello}}, \bibinfo
  {author} {\bibfnamefont {S.~D.}\ \bibnamefont {Conradson}}, \bibinfo {author}
  {\bibfnamefont {M.~G.}\ \bibnamefont {Samant}}, \ and\ \bibinfo {author}
  {\bibfnamefont {S.~S.~P.}\ \bibnamefont {Parkin}},\ }\href {\doibase
  10.1073/pnas.1419051112} {\bibfield  {journal} {\bibinfo  {journal}
  {Proceedings of the National Academy of Sciences}\ }\textbf {\bibinfo
  {volume} {112}},\ \bibinfo {pages} {1013} (\bibinfo {year}
  {2015})}\BibitemShut {NoStop}%
\bibitem [{\citenamefont {Okuyama}\ \emph {et~al.}(2014)\citenamefont
  {Okuyama}, \citenamefont {Nakano}, \citenamefont {Takeshita}, \citenamefont
  {Ohsumi}, \citenamefont {Tardif}, \citenamefont {Shibuya}, \citenamefont
  {Hatano}, \citenamefont {Yumoto}, \citenamefont {Koyama}, \citenamefont
  {Ohashi}, \citenamefont {Takata}, \citenamefont {Kawasaki}, \citenamefont
  {Arima}, \citenamefont {Tokura},\ and\ \citenamefont
  {Iwasa}}]{10.1063/1.4861901}%
  \BibitemOpen
  \bibfield  {author} {\bibinfo {author} {\bibfnamefont {D.}~\bibnamefont
  {Okuyama}}, \bibinfo {author} {\bibfnamefont {M.}~\bibnamefont {Nakano}},
  \bibinfo {author} {\bibfnamefont {S.}~\bibnamefont {Takeshita}}, \bibinfo
  {author} {\bibfnamefont {H.}~\bibnamefont {Ohsumi}}, \bibinfo {author}
  {\bibfnamefont {S.}~\bibnamefont {Tardif}}, \bibinfo {author} {\bibfnamefont
  {K.}~\bibnamefont {Shibuya}}, \bibinfo {author} {\bibfnamefont
  {T.}~\bibnamefont {Hatano}}, \bibinfo {author} {\bibfnamefont
  {H.}~\bibnamefont {Yumoto}}, \bibinfo {author} {\bibfnamefont
  {T.}~\bibnamefont {Koyama}}, \bibinfo {author} {\bibfnamefont
  {H.}~\bibnamefont {Ohashi}}, \bibinfo {author} {\bibfnamefont
  {M.}~\bibnamefont {Takata}}, \bibinfo {author} {\bibfnamefont
  {M.}~\bibnamefont {Kawasaki}}, \bibinfo {author} {\bibfnamefont
  {T.}~\bibnamefont {Arima}}, \bibinfo {author} {\bibfnamefont
  {Y.}~\bibnamefont {Tokura}}, \ and\ \bibinfo {author} {\bibfnamefont
  {Y.}~\bibnamefont {Iwasa}},\ }\href {\doibase 10.1063/1.4861901} {\bibfield
  {journal} {\bibinfo  {journal} {Applied Physics Letters}\ }\textbf {\bibinfo
  {volume} {104}},\ \bibinfo {pages} {023507} (\bibinfo {year}
  {2014})}\BibitemShut {NoStop}%
\bibitem [{\citenamefont {Dahlman}\ \emph {et~al.}(2016)\citenamefont
  {Dahlman}, \citenamefont {LeBlanc}, \citenamefont {Bergerud}, \citenamefont
  {Staller}, \citenamefont {Adair},\ and\ \citenamefont
  {Milliron}}]{10.1021/acs.nanolett.6b01756}%
  \BibitemOpen
  \bibfield  {author} {\bibinfo {author} {\bibfnamefont {C.~J.}\ \bibnamefont
  {Dahlman}}, \bibinfo {author} {\bibfnamefont {G.}~\bibnamefont {LeBlanc}},
  \bibinfo {author} {\bibfnamefont {A.}~\bibnamefont {Bergerud}}, \bibinfo
  {author} {\bibfnamefont {C.}~\bibnamefont {Staller}}, \bibinfo {author}
  {\bibfnamefont {J.}~\bibnamefont {Adair}}, \ and\ \bibinfo {author}
  {\bibfnamefont {D.~J.}\ \bibnamefont {Milliron}},\ }\href {\doibase
  10.1021/acs.nanolett.6b01756} {\bibfield  {journal} {\bibinfo  {journal}
  {Nano Letters}\ }\textbf {\bibinfo {volume} {16}},\ \bibinfo {pages} {6021}
  (\bibinfo {year} {2016})}\BibitemShut {NoStop}%
\bibitem [{\citenamefont {Yi}\ \emph {et~al.}(2014)\citenamefont {Yi},
  \citenamefont {Gao}, \citenamefont {Xie}, \citenamefont {Cheong},\ and\
  \citenamefont {Podzorov}}]{Yi2014}%
  \BibitemOpen
  \bibfield  {author} {\bibinfo {author} {\bibfnamefont {H.~T.}\ \bibnamefont
  {Yi}}, \bibinfo {author} {\bibfnamefont {B.}~\bibnamefont {Gao}}, \bibinfo
  {author} {\bibfnamefont {W.}~\bibnamefont {Xie}}, \bibinfo {author}
  {\bibfnamefont {S.-W.}\ \bibnamefont {Cheong}}, \ and\ \bibinfo {author}
  {\bibfnamefont {V.}~\bibnamefont {Podzorov}},\ }\href {\doibase
  10.1038/srep06604} {\bibfield  {journal} {\bibinfo  {journal} {Scientific
  Reports}\ }\textbf {\bibinfo {volume} {4}},\ \bibinfo {pages} {6604}
  (\bibinfo {year} {2014})}\BibitemShut {NoStop}%
\bibitem [{\citenamefont {Shimizu}\ \emph {et~al.}(2014)\citenamefont
  {Shimizu}, \citenamefont {Takahashi}, \citenamefont {Kubota}, \citenamefont
  {Kawasaki}, \citenamefont {Tokura},\ and\ \citenamefont
  {Iwasa}}]{10.1063/1.4899145}%
  \BibitemOpen
  \bibfield  {author} {\bibinfo {author} {\bibfnamefont {S.}~\bibnamefont
  {Shimizu}}, \bibinfo {author} {\bibfnamefont {K.~S.}\ \bibnamefont
  {Takahashi}}, \bibinfo {author} {\bibfnamefont {M.}~\bibnamefont {Kubota}},
  \bibinfo {author} {\bibfnamefont {M.}~\bibnamefont {Kawasaki}}, \bibinfo
  {author} {\bibfnamefont {Y.}~\bibnamefont {Tokura}}, \ and\ \bibinfo {author}
  {\bibfnamefont {Y.}~\bibnamefont {Iwasa}},\ }\href {\doibase
  10.1063/1.4899145} {\bibfield  {journal} {\bibinfo  {journal} {Applied
  Physics Letters}\ }\textbf {\bibinfo {volume} {105}},\ \bibinfo {pages}
  {163509} (\bibinfo {year} {2014})}\BibitemShut {NoStop}%
\bibitem [{\citenamefont {Shimizu}\ \emph {et~al.}(2015)\citenamefont
  {Shimizu}, \citenamefont {Ono}, \citenamefont {Hatano}, \citenamefont
  {Iwasa},\ and\ \citenamefont {Tokura}}]{PhysRevB.92.165304}%
  \BibitemOpen
  \bibfield  {author} {\bibinfo {author} {\bibfnamefont {S.}~\bibnamefont
  {Shimizu}}, \bibinfo {author} {\bibfnamefont {S.}~\bibnamefont {Ono}},
  \bibinfo {author} {\bibfnamefont {T.}~\bibnamefont {Hatano}}, \bibinfo
  {author} {\bibfnamefont {Y.}~\bibnamefont {Iwasa}}, \ and\ \bibinfo {author}
  {\bibfnamefont {Y.}~\bibnamefont {Tokura}},\ }\href {\doibase
  10.1103/PhysRevB.92.165304} {\bibfield  {journal} {\bibinfo  {journal} {Phys.
  Rev. B}\ }\textbf {\bibinfo {volume} {92}},\ \bibinfo {pages} {165304}
  (\bibinfo {year} {2015})}\BibitemShut {NoStop}%
\bibitem [{\citenamefont {Bollinger}\ \emph {et~al.}(2011)\citenamefont
  {Bollinger}, \citenamefont {Dubuis}, \citenamefont {Yoon}, \citenamefont
  {Pavuna}, \citenamefont {Misewich},\ and\ \citenamefont
  {Bozovic}}]{Bollinger2011}%
  \BibitemOpen
  \bibfield  {author} {\bibinfo {author} {\bibfnamefont {A.~T.}\ \bibnamefont
  {Bollinger}}, \bibinfo {author} {\bibfnamefont {G.}~\bibnamefont {Dubuis}},
  \bibinfo {author} {\bibfnamefont {J.}~\bibnamefont {Yoon}}, \bibinfo {author}
  {\bibfnamefont {D.}~\bibnamefont {Pavuna}}, \bibinfo {author} {\bibfnamefont
  {J.}~\bibnamefont {Misewich}}, \ and\ \bibinfo {author} {\bibfnamefont
  {I.}~\bibnamefont {Bozovic}},\ }\href {\doibase 10.1038/nature09998}
  {\bibfield  {journal} {\bibinfo  {journal} {Nature}\ }\textbf {\bibinfo
  {volume} {472}},\ \bibinfo {pages} {458} (\bibinfo {year}
  {2011})}\BibitemShut {NoStop}%
\bibitem [{\citenamefont {Leng}\ \emph {et~al.}(2011)\citenamefont {Leng},
  \citenamefont {Garcia-Barriocanal}, \citenamefont {Bose}, \citenamefont
  {Lee},\ and\ \citenamefont {Goldman}}]{PhysRevLett.107.027001}%
  \BibitemOpen
  \bibfield  {author} {\bibinfo {author} {\bibfnamefont {X.}~\bibnamefont
  {Leng}}, \bibinfo {author} {\bibfnamefont {J.}~\bibnamefont
  {Garcia-Barriocanal}}, \bibinfo {author} {\bibfnamefont {S.}~\bibnamefont
  {Bose}}, \bibinfo {author} {\bibfnamefont {Y.}~\bibnamefont {Lee}}, \ and\
  \bibinfo {author} {\bibfnamefont {A.~M.}\ \bibnamefont {Goldman}},\ }\href
  {\doibase 10.1103/PhysRevLett.107.027001} {\bibfield  {journal} {\bibinfo
  {journal} {Phys. Rev. Lett.}\ }\textbf {\bibinfo {volume} {107}},\ \bibinfo
  {pages} {027001} (\bibinfo {year} {2011})}\BibitemShut {NoStop}%
\bibitem [{\citenamefont {Ye}\ \emph {et~al.}(2012)\citenamefont {Ye},
  \citenamefont {Zhang}, \citenamefont {Akashi}, \citenamefont {Bahramy},
  \citenamefont {Arita},\ and\ \citenamefont {Iwasa}}]{Ye1193}%
  \BibitemOpen
  \bibfield  {author} {\bibinfo {author} {\bibfnamefont {J.~T.}\ \bibnamefont
  {Ye}}, \bibinfo {author} {\bibfnamefont {Y.~J.}\ \bibnamefont {Zhang}},
  \bibinfo {author} {\bibfnamefont {R.}~\bibnamefont {Akashi}}, \bibinfo
  {author} {\bibfnamefont {M.~S.}\ \bibnamefont {Bahramy}}, \bibinfo {author}
  {\bibfnamefont {R.}~\bibnamefont {Arita}}, \ and\ \bibinfo {author}
  {\bibfnamefont {Y.}~\bibnamefont {Iwasa}},\ }\href {\doibase
  10.1126/science.1228006} {\bibfield  {journal} {\bibinfo  {journal}
  {Science}\ }\textbf {\bibinfo {volume} {338}},\ \bibinfo {pages} {1193}
  (\bibinfo {year} {2012})}\BibitemShut {NoStop}%
\bibitem [{\citenamefont {Petach}\ \emph {et~al.}(2014)\citenamefont {Petach},
  \citenamefont {Lee}, \citenamefont {Davis}, \citenamefont {Mehta},\ and\
  \citenamefont {Goldhaber-Gordon}}]{Petach2014}%
  \BibitemOpen
  \bibfield  {author} {\bibinfo {author} {\bibfnamefont {T.~A.}\ \bibnamefont
  {Petach}}, \bibinfo {author} {\bibfnamefont {M.}~\bibnamefont {Lee}},
  \bibinfo {author} {\bibfnamefont {R.~C.}\ \bibnamefont {Davis}}, \bibinfo
  {author} {\bibfnamefont {A.}~\bibnamefont {Mehta}}, \ and\ \bibinfo {author}
  {\bibfnamefont {D.}~\bibnamefont {Goldhaber-Gordon}},\ }\href {\doibase
  10.1103/PhysRevB.90.081108} {\bibfield  {journal} {\bibinfo  {journal} {Phys.
  Rev. B}\ }\textbf {\bibinfo {volume} {90}},\ \bibinfo {pages} {081108}
  (\bibinfo {year} {2014})}\BibitemShut {NoStop}%
\bibitem [{\citenamefont {Bubel}\ \emph {et~al.}(2015)\citenamefont {Bubel},
  \citenamefont {Hauser}, \citenamefont {Glaudell}, \citenamefont {Mates},
  \citenamefont {Stemmer},\ and\ \citenamefont {Chabinyc}}]{Bubel2015}%
  \BibitemOpen
  \bibfield  {author} {\bibinfo {author} {\bibfnamefont {S.}~\bibnamefont
  {Bubel}}, \bibinfo {author} {\bibfnamefont {A.~J.}\ \bibnamefont {Hauser}},
  \bibinfo {author} {\bibfnamefont {A.~M.}\ \bibnamefont {Glaudell}}, \bibinfo
  {author} {\bibfnamefont {T.~E.}\ \bibnamefont {Mates}}, \bibinfo {author}
  {\bibfnamefont {S.}~\bibnamefont {Stemmer}}, \ and\ \bibinfo {author}
  {\bibfnamefont {M.~L.}\ \bibnamefont {Chabinyc}},\ }\href {\doibase
  10.1063/1.4915269} {\bibfield  {journal} {\bibinfo  {journal} {Applied
  Physics Letters}\ }\textbf {\bibinfo {volume} {106}},\ \bibinfo {pages}
  {122102} (\bibinfo {year} {2015})}\BibitemShut {NoStop}%
\bibitem [{\citenamefont {Leighton}(2019)}]{Leighton2019}%
  \BibitemOpen
  \bibfield  {author} {\bibinfo {author} {\bibfnamefont {C.}~\bibnamefont
  {Leighton}},\ }\href {\doibase 10.1038/s41563-018-0246-7} {\bibfield
  {journal} {\bibinfo  {journal} {Nature Materials}\ }\textbf {\bibinfo
  {volume} {18}},\ \bibinfo {pages} {13} (\bibinfo {year} {2019})}\BibitemShut
  {NoStop}%
\bibitem [{\citenamefont {Walter}\ \emph
  {et~al.}(2017{\natexlab{a}})\citenamefont {Walter}, \citenamefont {Yu},
  \citenamefont {Yu}, \citenamefont {Grutter}, \citenamefont {Kirby},
  \citenamefont {Borchers}, \citenamefont {Zhang}, \citenamefont {Zhou},
  \citenamefont {Birol}, \citenamefont {Greven},\ and\ \citenamefont
  {Leighton}}]{PhysRevMaterials.1.071403}%
  \BibitemOpen
  \bibfield  {author} {\bibinfo {author} {\bibfnamefont {J.}~\bibnamefont
  {Walter}}, \bibinfo {author} {\bibfnamefont {G.}~\bibnamefont {Yu}}, \bibinfo
  {author} {\bibfnamefont {B.}~\bibnamefont {Yu}}, \bibinfo {author}
  {\bibfnamefont {A.}~\bibnamefont {Grutter}}, \bibinfo {author} {\bibfnamefont
  {B.}~\bibnamefont {Kirby}}, \bibinfo {author} {\bibfnamefont
  {J.}~\bibnamefont {Borchers}}, \bibinfo {author} {\bibfnamefont
  {Z.}~\bibnamefont {Zhang}}, \bibinfo {author} {\bibfnamefont
  {H.}~\bibnamefont {Zhou}}, \bibinfo {author} {\bibfnamefont {T.}~\bibnamefont
  {Birol}}, \bibinfo {author} {\bibfnamefont {M.}~\bibnamefont {Greven}}, \
  and\ \bibinfo {author} {\bibfnamefont {C.}~\bibnamefont {Leighton}},\ }\href
  {\doibase 10.1103/PhysRevMaterials.1.071403} {\bibfield  {journal} {\bibinfo
  {journal} {Phys. Rev. Materials}\ }\textbf {\bibinfo {volume} {1}},\ \bibinfo
  {pages} {071403} (\bibinfo {year} {2017}{\natexlab{a}})}\BibitemShut
  {NoStop}%
\bibitem [{\citenamefont {Chen}\ \emph {et~al.}(2009)\citenamefont {Chen},
  \citenamefont {Qing}, \citenamefont {Xia}, \citenamefont {Li},\ and\
  \citenamefont {Tao}}]{Chen2009}%
  \BibitemOpen
  \bibfield  {author} {\bibinfo {author} {\bibfnamefont {F.}~\bibnamefont
  {Chen}}, \bibinfo {author} {\bibfnamefont {Q.}~\bibnamefont {Qing}}, \bibinfo
  {author} {\bibfnamefont {J.}~\bibnamefont {Xia}}, \bibinfo {author}
  {\bibfnamefont {J.}~\bibnamefont {Li}}, \ and\ \bibinfo {author}
  {\bibfnamefont {N.}~\bibnamefont {Tao}},\ }\href {\doibase 10.1021/ja9041862}
  {\bibfield  {journal} {\bibinfo  {journal} {Journal of the American Chemical
  Society}\ }\textbf {\bibinfo {volume} {131}},\ \bibinfo {pages} {9908}
  (\bibinfo {year} {2009})}\BibitemShut {NoStop}%
\bibitem [{\citenamefont {Yuan}\ \emph {et~al.}(2010)\citenamefont {Yuan},
  \citenamefont {Shimotani}, \citenamefont {Ye}, \citenamefont {Yoon},
  \citenamefont {Aliah}, \citenamefont {Tsukazaki}, \citenamefont {Kawasaki},\
  and\ \citenamefont {Iwasa}}]{Yuan2010}%
  \BibitemOpen
  \bibfield  {author} {\bibinfo {author} {\bibfnamefont {H.}~\bibnamefont
  {Yuan}}, \bibinfo {author} {\bibfnamefont {H.}~\bibnamefont {Shimotani}},
  \bibinfo {author} {\bibfnamefont {J.}~\bibnamefont {Ye}}, \bibinfo {author}
  {\bibfnamefont {S.}~\bibnamefont {Yoon}}, \bibinfo {author} {\bibfnamefont
  {H.}~\bibnamefont {Aliah}}, \bibinfo {author} {\bibfnamefont
  {A.}~\bibnamefont {Tsukazaki}}, \bibinfo {author} {\bibfnamefont
  {M.}~\bibnamefont {Kawasaki}}, \ and\ \bibinfo {author} {\bibfnamefont
  {Y.}~\bibnamefont {Iwasa}},\ }\href {\doibase 10.1021/ja108912x} {\bibfield
  {journal} {\bibinfo  {journal} {Journal of the American Chemical Society}\
  }\textbf {\bibinfo {volume} {132}},\ \bibinfo {pages} {18402} (\bibinfo
  {year} {2010})}\BibitemShut {NoStop}%
\bibitem [{\citenamefont {Kay}\ \emph {et~al.}(2012)\citenamefont {Kay},
  \citenamefont {Higgins}, \citenamefont {Jeppesen}, \citenamefont {Leary},
  \citenamefont {Lycoops}, \citenamefont {Ulstrup},\ and\ \citenamefont
  {Nichols}}]{Kay2012}%
  \BibitemOpen
  \bibfield  {author} {\bibinfo {author} {\bibfnamefont {N.~J.}\ \bibnamefont
  {Kay}}, \bibinfo {author} {\bibfnamefont {S.~J.}\ \bibnamefont {Higgins}},
  \bibinfo {author} {\bibfnamefont {J.~O.}\ \bibnamefont {Jeppesen}}, \bibinfo
  {author} {\bibfnamefont {E.}~\bibnamefont {Leary}}, \bibinfo {author}
  {\bibfnamefont {J.}~\bibnamefont {Lycoops}}, \bibinfo {author} {\bibfnamefont
  {J.}~\bibnamefont {Ulstrup}}, \ and\ \bibinfo {author} {\bibfnamefont
  {R.~J.}\ \bibnamefont {Nichols}},\ }\href {\doibase 10.1021/ja307407e}
  {\bibfield  {journal} {\bibinfo  {journal} {Journal of the American Chemical
  Society}\ }\textbf {\bibinfo {volume} {134}},\ \bibinfo {pages} {16817}
  (\bibinfo {year} {2012})}\BibitemShut {NoStop}%
\bibitem [{\citenamefont {Ge}\ \emph {et~al.}(2015)\citenamefont {Ge},
  \citenamefont {Jin}, \citenamefont {Gu}, \citenamefont {Peng}, \citenamefont
  {Hu}, \citenamefont {Guo}, \citenamefont {Shi}, \citenamefont {Li},
  \citenamefont {Wang}, \citenamefont {Guo}, \citenamefont {Wang},
  \citenamefont {He}, \citenamefont {Lu},\ and\ \citenamefont {Yang}}]{Ge2015}%
  \BibitemOpen
  \bibfield  {author} {\bibinfo {author} {\bibfnamefont {C.}~\bibnamefont
  {Ge}}, \bibinfo {author} {\bibfnamefont {K.-J.}\ \bibnamefont {Jin}},
  \bibinfo {author} {\bibfnamefont {L.}~\bibnamefont {Gu}}, \bibinfo {author}
  {\bibfnamefont {L.-C.}\ \bibnamefont {Peng}}, \bibinfo {author}
  {\bibfnamefont {Y.-S.}\ \bibnamefont {Hu}}, \bibinfo {author} {\bibfnamefont
  {H.-Z.}\ \bibnamefont {Guo}}, \bibinfo {author} {\bibfnamefont {H.-F.}\
  \bibnamefont {Shi}}, \bibinfo {author} {\bibfnamefont {J.-K.}\ \bibnamefont
  {Li}}, \bibinfo {author} {\bibfnamefont {J.-O.}\ \bibnamefont {Wang}},
  \bibinfo {author} {\bibfnamefont {X.-X.}\ \bibnamefont {Guo}}, \bibinfo
  {author} {\bibfnamefont {C.}~\bibnamefont {Wang}}, \bibinfo {author}
  {\bibfnamefont {M.}~\bibnamefont {He}}, \bibinfo {author} {\bibfnamefont
  {H.-B.}\ \bibnamefont {Lu}}, \ and\ \bibinfo {author} {\bibfnamefont {G.-Z.}\
  \bibnamefont {Yang}},\ }\href {\doibase 10.1002/admi.201500407} {\bibfield
  {journal} {\bibinfo  {journal} {Advanced Materials Interfaces}\ }\textbf
  {\bibinfo {volume} {2}},\ \bibinfo {pages} {1500407} (\bibinfo {year}
  {2015})}\BibitemShut {NoStop}%
\bibitem [{\citenamefont {Ramesha}\ \emph {et~al.}(2004)\citenamefont
  {Ramesha}, \citenamefont {Seshadri}, \citenamefont {Ederer}, \citenamefont
  {He},\ and\ \citenamefont {Subramanian}}]{ramesha_experimental_2004}%
  \BibitemOpen
  \bibfield  {author} {\bibinfo {author} {\bibfnamefont {K.}~\bibnamefont
  {Ramesha}}, \bibinfo {author} {\bibfnamefont {R.}~\bibnamefont {Seshadri}},
  \bibinfo {author} {\bibfnamefont {C.}~\bibnamefont {Ederer}}, \bibinfo
  {author} {\bibfnamefont {T.}~\bibnamefont {He}}, \ and\ \bibinfo {author}
  {\bibfnamefont {M.~A.}\ \bibnamefont {Subramanian}},\ }\href {\doibase
  10.1103/PhysRevB.70.214409} {\bibfield  {journal} {\bibinfo  {journal} {Phys.
  Rev. B}\ }\textbf {\bibinfo {volume} {70}},\ \bibinfo {pages} {214409}
  (\bibinfo {year} {2004})}\BibitemShut {NoStop}%
\bibitem [{\citenamefont {Sun}\ \emph {et~al.}(2011)\citenamefont {Sun},
  \citenamefont {Chan}, \citenamefont {Kang},\ and\ \citenamefont
  {Ceder}}]{Sun2011}%
  \BibitemOpen
  \bibfield  {author} {\bibinfo {author} {\bibfnamefont {R.}~\bibnamefont
  {Sun}}, \bibinfo {author} {\bibfnamefont {M.~K.~Y.}\ \bibnamefont {Chan}},
  \bibinfo {author} {\bibfnamefont {S.~Y.}\ \bibnamefont {Kang}}, \ and\
  \bibinfo {author} {\bibfnamefont {G.}~\bibnamefont {Ceder}},\ }\href
  {\doibase 10.1103/PhysRevB.84.035212} {\bibfield  {journal} {\bibinfo
  {journal} {Physical Review B}\ }\textbf {\bibinfo {volume} {84}},\ \bibinfo
  {pages} {035212} (\bibinfo {year} {2011})}\BibitemShut {NoStop}%
\bibitem [{\citenamefont {Ray}\ \emph {et~al.}()\citenamefont {Ray},
  \citenamefont {Voigt}, \citenamefont {Manno}, \citenamefont {Leighton},
  \citenamefont {Aydil},\ and\ \citenamefont {Gagliardi}}]{Gagliardi}%
  \BibitemOpen
  \bibfield  {author} {\bibinfo {author} {\bibnamefont {Ray}}, \bibinfo
  {author} {\bibnamefont {Voigt}}, \bibinfo {author} {\bibnamefont {Manno}},
  \bibinfo {author} {\bibnamefont {Leighton}}, \bibinfo {author} {\bibnamefont
  {Aydil}}, \ and\ \bibinfo {author} {\bibnamefont {Gagliardi}},\ }\href@noop
  {} {\emph {\bibinfo {title} {Under Review.}}}\BibitemShut {Stop}%
\bibitem [{\citenamefont {Altermatt}\ \emph {et~al.}(2002)\citenamefont
  {Altermatt}, \citenamefont {Kiesewetter}, \citenamefont {Ellmer},\ and\
  \citenamefont {Tributsch}}]{ALTERMATT2002181}%
  \BibitemOpen
  \bibfield  {author} {\bibinfo {author} {\bibfnamefont {P.~P.}\ \bibnamefont
  {Altermatt}}, \bibinfo {author} {\bibfnamefont {T.}~\bibnamefont
  {Kiesewetter}}, \bibinfo {author} {\bibfnamefont {K.}~\bibnamefont {Ellmer}},
  \ and\ \bibinfo {author} {\bibfnamefont {H.}~\bibnamefont {Tributsch}},\
  }\href {\doibase https://doi.org/10.1016/S0927-0248(01)00053-8} {\bibfield
  {journal} {\bibinfo  {journal} {Solar Energy Materials and Solar Cells}\
  }\textbf {\bibinfo {volume} {71}},\ \bibinfo {pages} {181 } (\bibinfo {year}
  {2002})}\BibitemShut {NoStop}%
\bibitem [{\citenamefont {Bi}\ \emph {et~al.}(2011)\citenamefont {Bi},
  \citenamefont {Yuan}, \citenamefont {Exstrom}, \citenamefont {Darveau},\ and\
  \citenamefont {Huang}}]{10.1021/nl202902z}%
  \BibitemOpen
  \bibfield  {author} {\bibinfo {author} {\bibfnamefont {Y.}~\bibnamefont
  {Bi}}, \bibinfo {author} {\bibfnamefont {Y.}~\bibnamefont {Yuan}}, \bibinfo
  {author} {\bibfnamefont {C.~L.}\ \bibnamefont {Exstrom}}, \bibinfo {author}
  {\bibfnamefont {S.~A.}\ \bibnamefont {Darveau}}, \ and\ \bibinfo {author}
  {\bibfnamefont {J.}~\bibnamefont {Huang}},\ }\href {\doibase
  10.1021/nl202902z} {\bibfield  {journal} {\bibinfo  {journal} {Nano Letters}\
  }\textbf {\bibinfo {volume} {11}},\ \bibinfo {pages} {4953} (\bibinfo {year}
  {2011})}\BibitemShut {NoStop}%
\bibitem [{\citenamefont {Alharbi}\ \emph {et~al.}(2011)\citenamefont
  {Alharbi}, \citenamefont {Bass}, \citenamefont {Salhi}, \citenamefont
  {Alyamani}, \citenamefont {Kim},\ and\ \citenamefont
  {Miller}}]{ALHARBI20112753}%
  \BibitemOpen
  \bibfield  {author} {\bibinfo {author} {\bibfnamefont {F.}~\bibnamefont
  {Alharbi}}, \bibinfo {author} {\bibfnamefont {J.~D.}\ \bibnamefont {Bass}},
  \bibinfo {author} {\bibfnamefont {A.}~\bibnamefont {Salhi}}, \bibinfo
  {author} {\bibfnamefont {A.}~\bibnamefont {Alyamani}}, \bibinfo {author}
  {\bibfnamefont {H.-C.}\ \bibnamefont {Kim}}, \ and\ \bibinfo {author}
  {\bibfnamefont {R.~D.}\ \bibnamefont {Miller}},\ }\href {\doibase
  https://doi.org/10.1016/j.renene.2011.03.010} {\bibfield  {journal} {\bibinfo
   {journal} {Renewable Energy}\ }\textbf {\bibinfo {volume} {36}},\ \bibinfo
  {pages} {2753 } (\bibinfo {year} {2011})}\BibitemShut {NoStop}%
\bibitem [{\citenamefont {Umemoto}\ \emph {et~al.}(2006)\citenamefont
  {Umemoto}, \citenamefont {Wentzcovitch}, \citenamefont {Wang},\ and\
  \citenamefont {Leighton}}]{10.1002/pssb.200666821}%
  \BibitemOpen
  \bibfield  {author} {\bibinfo {author} {\bibfnamefont {K.}~\bibnamefont
  {Umemoto}}, \bibinfo {author} {\bibfnamefont {R.~M.}\ \bibnamefont
  {Wentzcovitch}}, \bibinfo {author} {\bibfnamefont {L.}~\bibnamefont {Wang}},
  \ and\ \bibinfo {author} {\bibfnamefont {C.}~\bibnamefont {Leighton}},\
  }\href {\doibase 10.1002/pssb.200666821} {\bibfield  {journal} {\bibinfo
  {journal} {physica status solidi (b)}\ }\textbf {\bibinfo {volume} {243}},\
  \bibinfo {pages} {2117} (\bibinfo {year} {2006})}\BibitemShut {NoStop}%
\bibitem [{\citenamefont {Voigt}\ \emph {et~al.}(2019)\citenamefont {Voigt},
  \citenamefont {Moore}, \citenamefont {Manno}, \citenamefont {Walter},
  \citenamefont {Jeremiason}, \citenamefont {Aydil},\ and\ \citenamefont
  {Leighton}}]{Voigt2019}%
  \BibitemOpen
  \bibfield  {author} {\bibinfo {author} {\bibfnamefont {B.}~\bibnamefont
  {Voigt}}, \bibinfo {author} {\bibfnamefont {W.}~\bibnamefont {Moore}},
  \bibinfo {author} {\bibfnamefont {M.}~\bibnamefont {Manno}}, \bibinfo
  {author} {\bibfnamefont {J.}~\bibnamefont {Walter}}, \bibinfo {author}
  {\bibfnamefont {J.~D.}\ \bibnamefont {Jeremiason}}, \bibinfo {author}
  {\bibfnamefont {E.~S.}\ \bibnamefont {Aydil}}, \ and\ \bibinfo {author}
  {\bibfnamefont {C.}~\bibnamefont {Leighton}},\ }\href {\doibase
  10.1021/acsami.9b01335} {\bibfield  {journal} {\bibinfo  {journal} {ACS
  Applied Materials and Interfaces}\ }\textbf {\bibinfo {volume} {11}},\
  \bibinfo {pages} {15552} (\bibinfo {year} {2019})}\BibitemShut {NoStop}%
\bibitem [{\citenamefont {Limpinsel}\ \emph {et~al.}(2014)\citenamefont
  {Limpinsel}, \citenamefont {Farhi}, \citenamefont {Berry}, \citenamefont
  {Lindemuth}, \citenamefont {Perkins}, \citenamefont {Lin},\ and\
  \citenamefont {Law}}]{Limpinsel2014}%
  \BibitemOpen
  \bibfield  {author} {\bibinfo {author} {\bibfnamefont {M.}~\bibnamefont
  {Limpinsel}}, \bibinfo {author} {\bibfnamefont {N.}~\bibnamefont {Farhi}},
  \bibinfo {author} {\bibfnamefont {N.}~\bibnamefont {Berry}}, \bibinfo
  {author} {\bibfnamefont {J.}~\bibnamefont {Lindemuth}}, \bibinfo {author}
  {\bibfnamefont {C.~L.}\ \bibnamefont {Perkins}}, \bibinfo {author}
  {\bibfnamefont {Q.}~\bibnamefont {Lin}}, \ and\ \bibinfo {author}
  {\bibfnamefont {M.}~\bibnamefont {Law}},\ }\href {\doibase
  10.1039/c3ee43169j} {\bibfield  {journal} {\bibinfo  {journal} {Energy and
  Environmental Science}\ }\textbf {\bibinfo {volume} {7}},\ \bibinfo {pages}
  {1974} (\bibinfo {year} {2014})}\BibitemShut {NoStop}%
\bibitem [{\citenamefont {Liang}\ \emph {et~al.}(2014)\citenamefont {Liang},
  \citenamefont {Cab{\'{a}}n-Acevedo}, \citenamefont {Kaiser},\ and\
  \citenamefont {Jin}}]{Liang2014}%
  \BibitemOpen
  \bibfield  {author} {\bibinfo {author} {\bibfnamefont {D.}~\bibnamefont
  {Liang}}, \bibinfo {author} {\bibfnamefont {M.}~\bibnamefont
  {Cab{\'{a}}n-Acevedo}}, \bibinfo {author} {\bibfnamefont {N.~S.}\
  \bibnamefont {Kaiser}}, \ and\ \bibinfo {author} {\bibfnamefont
  {S.}~\bibnamefont {Jin}},\ }\href {\doibase 10.1021/nl501942w} {\bibfield
  {journal} {\bibinfo  {journal} {Nano Letters}\ }\textbf {\bibinfo {volume}
  {14}},\ \bibinfo {pages} {6754} (\bibinfo {year} {2014})}\BibitemShut
  {NoStop}%
\bibitem [{\citenamefont {Zhang}\ \emph {et~al.}(2017)\citenamefont {Zhang},
  \citenamefont {Li}, \citenamefont {Walter}, \citenamefont {O'Brien},
  \citenamefont {Manno}, \citenamefont {Voigt}, \citenamefont {Mork},
  \citenamefont {Baryshev}, \citenamefont {Kakalios}, \citenamefont {Aydil},\
  and\ \citenamefont {Leighton}}]{Zhang2017}%
  \BibitemOpen
  \bibfield  {author} {\bibinfo {author} {\bibfnamefont {X.}~\bibnamefont
  {Zhang}}, \bibinfo {author} {\bibfnamefont {M.}~\bibnamefont {Li}}, \bibinfo
  {author} {\bibfnamefont {J.}~\bibnamefont {Walter}}, \bibinfo {author}
  {\bibfnamefont {L.}~\bibnamefont {O'Brien}}, \bibinfo {author} {\bibfnamefont
  {M.~A.}\ \bibnamefont {Manno}}, \bibinfo {author} {\bibfnamefont
  {B.}~\bibnamefont {Voigt}}, \bibinfo {author} {\bibfnamefont
  {F.}~\bibnamefont {Mork}}, \bibinfo {author} {\bibfnamefont {S.~V.}\
  \bibnamefont {Baryshev}}, \bibinfo {author} {\bibfnamefont {J.}~\bibnamefont
  {Kakalios}}, \bibinfo {author} {\bibfnamefont {E.~S.}\ \bibnamefont {Aydil}},
  \ and\ \bibinfo {author} {\bibfnamefont {C.}~\bibnamefont {Leighton}},\
  }\href {\doibase 10.1103/PhysRevMaterials.1.015402} {\bibfield  {journal}
  {\bibinfo  {journal} {Phys. Rev. Materials}\ }\textbf {\bibinfo {volume}
  {1}},\ \bibinfo {pages} {015402} (\bibinfo {year} {2017})}\BibitemShut
  {NoStop}%
\bibitem [{\citenamefont {Walter}\ \emph
  {et~al.}(2017{\natexlab{b}})\citenamefont {Walter}, \citenamefont {Zhang},
  \citenamefont {Voigt}, \citenamefont {Hool}, \citenamefont {Manno},
  \citenamefont {Mork}, \citenamefont {Aydil},\ and\ \citenamefont
  {Leighton}}]{walter_surface_2017}%
  \BibitemOpen
  \bibfield  {author} {\bibinfo {author} {\bibfnamefont {J.}~\bibnamefont
  {Walter}}, \bibinfo {author} {\bibfnamefont {X.}~\bibnamefont {Zhang}},
  \bibinfo {author} {\bibfnamefont {B.}~\bibnamefont {Voigt}}, \bibinfo
  {author} {\bibfnamefont {R.}~\bibnamefont {Hool}}, \bibinfo {author}
  {\bibfnamefont {M.}~\bibnamefont {Manno}}, \bibinfo {author} {\bibfnamefont
  {F.}~\bibnamefont {Mork}}, \bibinfo {author} {\bibfnamefont {E.~S.}\
  \bibnamefont {Aydil}}, \ and\ \bibinfo {author} {\bibfnamefont
  {C.}~\bibnamefont {Leighton}},\ }\href {\doibase
  10.1103/PhysRevMaterials.1.065403} {\bibfield  {journal} {\bibinfo  {journal}
  {Phys. Rev. Materials}\ }\textbf {\bibinfo {volume} {1}},\ \bibinfo {pages}
  {065403} (\bibinfo {year} {2017}{\natexlab{b}})}\BibitemShut {NoStop}%
\bibitem [{\citenamefont {Guo}\ \emph {et~al.}(2008)\citenamefont {Guo},
  \citenamefont {Young}, \citenamefont {Macaluso}, \citenamefont {Browne},
  \citenamefont {Henderson}, \citenamefont {Chan}, \citenamefont {Henry},\ and\
  \citenamefont {DiTusa}}]{guo_discovery_2008}%
  \BibitemOpen
  \bibfield  {author} {\bibinfo {author} {\bibfnamefont {S.}~\bibnamefont
  {Guo}}, \bibinfo {author} {\bibfnamefont {D.~P.}\ \bibnamefont {Young}},
  \bibinfo {author} {\bibfnamefont {R.~T.}\ \bibnamefont {Macaluso}}, \bibinfo
  {author} {\bibfnamefont {D.~A.}\ \bibnamefont {Browne}}, \bibinfo {author}
  {\bibfnamefont {N.~L.}\ \bibnamefont {Henderson}}, \bibinfo {author}
  {\bibfnamefont {J.~Y.}\ \bibnamefont {Chan}}, \bibinfo {author}
  {\bibfnamefont {L.~L.}\ \bibnamefont {Henry}}, \ and\ \bibinfo {author}
  {\bibfnamefont {J.~F.}\ \bibnamefont {DiTusa}},\ }\href {\doibase
  10.1103/PhysRevLett.100.017209} {\bibfield  {journal} {\bibinfo  {journal}
  {Phys. Rev. Lett.}\ }\textbf {\bibinfo {volume} {100}},\ \bibinfo {pages}
  {017209} (\bibinfo {year} {2008})}\BibitemShut {NoStop}%
\bibitem [{\citenamefont {Guo}\ \emph {et~al.}(2010)\citenamefont {Guo},
  \citenamefont {Young}, \citenamefont {Macaluso}, \citenamefont {Browne},
  \citenamefont {Henderson}, \citenamefont {Chan}, \citenamefont {Henry},\ and\
  \citenamefont {DiTusa}}]{PhysRevB.81.144423}%
  \BibitemOpen
  \bibfield  {author} {\bibinfo {author} {\bibfnamefont {S.}~\bibnamefont
  {Guo}}, \bibinfo {author} {\bibfnamefont {D.~P.}\ \bibnamefont {Young}},
  \bibinfo {author} {\bibfnamefont {R.~T.}\ \bibnamefont {Macaluso}}, \bibinfo
  {author} {\bibfnamefont {D.~A.}\ \bibnamefont {Browne}}, \bibinfo {author}
  {\bibfnamefont {N.~L.}\ \bibnamefont {Henderson}}, \bibinfo {author}
  {\bibfnamefont {J.~Y.}\ \bibnamefont {Chan}}, \bibinfo {author}
  {\bibfnamefont {L.~L.}\ \bibnamefont {Henry}}, \ and\ \bibinfo {author}
  {\bibfnamefont {J.~F.}\ \bibnamefont {DiTusa}},\ }\href {\doibase
  10.1103/PhysRevB.81.144423} {\bibfield  {journal} {\bibinfo  {journal} {Phys.
  Rev. B}\ }\textbf {\bibinfo {volume} {81}},\ \bibinfo {pages} {144423}
  (\bibinfo {year} {2010})}\BibitemShut {NoStop}%
\bibitem [{\citenamefont {Kotiuga}\ and\ \citenamefont
  {Rabe}(2019)}]{kotiuga2019highdensity}%
  \BibitemOpen
  \bibfield  {author} {\bibinfo {author} {\bibfnamefont {M.}~\bibnamefont
  {Kotiuga}}\ and\ \bibinfo {author} {\bibfnamefont {K.~M.}\ \bibnamefont
  {Rabe}},\ }\href@noop {} {\enquote {\bibinfo {title} {High-density electron
  doping of smnio$_3$ from first principles},}\ } (\bibinfo {year} {2019}),\
  \Eprint {http://arxiv.org/abs/1909.03425} {arXiv:1909.03425
  [cond-mat.mes-hall]} \BibitemShut {NoStop}%
\bibitem [{\citenamefont {Kresse}\ and\ \citenamefont
  {Furthm\"uller}(1996)}]{VASP}%
  \BibitemOpen
  \bibfield  {author} {\bibinfo {author} {\bibfnamefont {G.}~\bibnamefont
  {Kresse}}\ and\ \bibinfo {author} {\bibfnamefont {J.}~\bibnamefont
  {Furthm\"uller}},\ }\href {\doibase 10.1103/PhysRevB.54.11169} {\bibfield
  {journal} {\bibinfo  {journal} {Phys. Rev. B}\ }\textbf {\bibinfo {volume}
  {54}},\ \bibinfo {pages} {11169} (\bibinfo {year} {1996})}\BibitemShut
  {NoStop}%
\bibitem [{\citenamefont {Kresse}\ and\ \citenamefont
  {Joubert}(1999)}]{VASP-PAW}%
  \BibitemOpen
  \bibfield  {author} {\bibinfo {author} {\bibfnamefont {G.}~\bibnamefont
  {Kresse}}\ and\ \bibinfo {author} {\bibfnamefont {D.}~\bibnamefont
  {Joubert}},\ }\href {\doibase 10.1103/PhysRevB.59.1758} {\bibfield  {journal}
  {\bibinfo  {journal} {Phys. Rev. B}\ }\textbf {\bibinfo {volume} {59}},\
  \bibinfo {pages} {1758} (\bibinfo {year} {1999})}\BibitemShut {NoStop}%
\bibitem [{\citenamefont {Perdew}\ \emph {et~al.}(2008)\citenamefont {Perdew},
  \citenamefont {Ruzsinszky}, \citenamefont {Csonka}, \citenamefont {Vydrov},
  \citenamefont {Scuseria}, \citenamefont {Constantin}, \citenamefont {Zhou},\
  and\ \citenamefont {Burke}}]{PBEsol}%
  \BibitemOpen
  \bibfield  {author} {\bibinfo {author} {\bibfnamefont {J.~P.}\ \bibnamefont
  {Perdew}}, \bibinfo {author} {\bibfnamefont {A.}~\bibnamefont {Ruzsinszky}},
  \bibinfo {author} {\bibfnamefont {G.~I.}\ \bibnamefont {Csonka}}, \bibinfo
  {author} {\bibfnamefont {O.~A.}\ \bibnamefont {Vydrov}}, \bibinfo {author}
  {\bibfnamefont {G.~E.}\ \bibnamefont {Scuseria}}, \bibinfo {author}
  {\bibfnamefont {L.~A.}\ \bibnamefont {Constantin}}, \bibinfo {author}
  {\bibfnamefont {X.}~\bibnamefont {Zhou}}, \ and\ \bibinfo {author}
  {\bibfnamefont {K.}~\bibnamefont {Burke}},\ }\href {\doibase
  10.1103/PhysRevLett.100.136406} {\bibfield  {journal} {\bibinfo  {journal}
  {Phys. Rev. Lett.}\ }\textbf {\bibinfo {volume} {100}},\ \bibinfo {pages}
  {136406} (\bibinfo {year} {2008})}\BibitemShut {NoStop}%
\bibitem [{\citenamefont {Dudarev}\ \emph {et~al.}(1998)\citenamefont
  {Dudarev}, \citenamefont {Botton}, \citenamefont {Savrasov}, \citenamefont
  {Humphreys},\ and\ \citenamefont {Sutton}}]{Dudarev1998}%
  \BibitemOpen
  \bibfield  {author} {\bibinfo {author} {\bibfnamefont {S.~L.}\ \bibnamefont
  {Dudarev}}, \bibinfo {author} {\bibfnamefont {G.~A.}\ \bibnamefont {Botton}},
  \bibinfo {author} {\bibfnamefont {S.~Y.}\ \bibnamefont {Savrasov}}, \bibinfo
  {author} {\bibfnamefont {C.~J.}\ \bibnamefont {Humphreys}}, \ and\ \bibinfo
  {author} {\bibfnamefont {A.~P.}\ \bibnamefont {Sutton}},\ }\href {\doibase
  10.1103/PhysRevB.57.1505} {\bibfield  {journal} {\bibinfo  {journal} {Phys.
  Rev. B}\ }\textbf {\bibinfo {volume} {57}},\ \bibinfo {pages} {1505}
  (\bibinfo {year} {1998})}\BibitemShut {NoStop}%
\bibitem [{\citenamefont {Hu}\ \emph {et~al.}(2012)\citenamefont {Hu},
  \citenamefont {Zhang}, \citenamefont {Law},\ and\ \citenamefont
  {Wu}}]{hu_first-principles_2012}%
  \BibitemOpen
  \bibfield  {author} {\bibinfo {author} {\bibfnamefont {J.}~\bibnamefont
  {Hu}}, \bibinfo {author} {\bibfnamefont {Y.}~\bibnamefont {Zhang}}, \bibinfo
  {author} {\bibfnamefont {M.}~\bibnamefont {Law}}, \ and\ \bibinfo {author}
  {\bibfnamefont {R.}~\bibnamefont {Wu}},\ }\href {\doibase
  10.1103/PhysRevB.85.085203} {\bibfield  {journal} {\bibinfo  {journal} {Phys.
  Rev. B}\ }\textbf {\bibinfo {volume} {85}},\ \bibinfo {pages} {085203}
  (\bibinfo {year} {2012})}\BibitemShut {NoStop}%
\bibitem [{\citenamefont {Mazin}(2000)}]{mazin_robust_2000}%
  \BibitemOpen
  \bibfield  {author} {\bibinfo {author} {\bibfnamefont {I.~I.}\ \bibnamefont
  {Mazin}},\ }\href {\doibase 10.1063/1.1324720} {\bibfield  {journal}
  {\bibinfo  {journal} {Applied Physics Letters}\ }\textbf {\bibinfo {volume}
  {77}},\ \bibinfo {pages} {3000} (\bibinfo {year} {2000})}\BibitemShut
  {NoStop}%
\bibitem [{\citenamefont {Feng}\ \emph {et~al.}(2018)\citenamefont {Feng},
  \citenamefont {Yang},\ and\ \citenamefont {Zhang}}]{feng_structural_2018}%
  \BibitemOpen
  \bibfield  {author} {\bibinfo {author} {\bibfnamefont {Z.-Y.}\ \bibnamefont
  {Feng}}, \bibinfo {author} {\bibfnamefont {Y.}~\bibnamefont {Yang}}, \ and\
  \bibinfo {author} {\bibfnamefont {J.-M.}\ \bibnamefont {Zhang}},\ }\href
  {\doibase 10.1088/2053-1591/aaa362} {\bibfield  {journal} {\bibinfo
  {journal} {Materials Research Express}\ }\textbf {\bibinfo {volume} {5}},\
  \bibinfo {pages} {016507} (\bibinfo {year} {2018})}\BibitemShut {NoStop}%
\bibitem [{\citenamefont {Mostofi}\ \emph {et~al.}(2014)\citenamefont
  {Mostofi}, \citenamefont {Yates}, \citenamefont {Pizzi}, \citenamefont {Lee},
  \citenamefont {Souza}, \citenamefont {Vanderbilt},\ and\ \citenamefont
  {Marzari}}]{wannier90}%
  \BibitemOpen
  \bibfield  {author} {\bibinfo {author} {\bibfnamefont {A.~A.}\ \bibnamefont
  {Mostofi}}, \bibinfo {author} {\bibfnamefont {J.~R.}\ \bibnamefont {Yates}},
  \bibinfo {author} {\bibfnamefont {G.}~\bibnamefont {Pizzi}}, \bibinfo
  {author} {\bibfnamefont {Y.-S.}\ \bibnamefont {Lee}}, \bibinfo {author}
  {\bibfnamefont {I.}~\bibnamefont {Souza}}, \bibinfo {author} {\bibfnamefont
  {D.}~\bibnamefont {Vanderbilt}}, \ and\ \bibinfo {author} {\bibfnamefont
  {N.}~\bibnamefont {Marzari}},\ }\href {\doibase
  https://doi.org/10.1016/j.cpc.2014.05.003} {\bibfield  {journal} {\bibinfo
  {journal} {Computer Physics Communications}\ }\textbf {\bibinfo {volume}
  {185}},\ \bibinfo {pages} {2309 } (\bibinfo {year} {2014})}\BibitemShut
  {NoStop}%
\bibitem [{Note1()}]{Note1}%
  \BibitemOpen
  \bibinfo {note} {As is standard in similar DFT calculations, a homogeneous
  background charge is also added to ensure charge neutrality of the unit
  cell.}\BibitemShut {Stop}%
\bibitem [{\citenamefont {Souza}\ \emph {et~al.}(2001)\citenamefont {Souza},
  \citenamefont {Marzari},\ and\ \citenamefont
  {Vanderbilt}}]{wannier_entangled}%
  \BibitemOpen
  \bibfield  {author} {\bibinfo {author} {\bibfnamefont {I.}~\bibnamefont
  {Souza}}, \bibinfo {author} {\bibfnamefont {N.}~\bibnamefont {Marzari}}, \
  and\ \bibinfo {author} {\bibfnamefont {D.}~\bibnamefont {Vanderbilt}},\
  }\href {\doibase 10.1103/PhysRevB.65.035109} {\bibfield  {journal} {\bibinfo
  {journal} {Phys. Rev. B}\ }\textbf {\bibinfo {volume} {65}},\ \bibinfo
  {pages} {035109} (\bibinfo {year} {2001})}\BibitemShut {NoStop}%
\bibitem [{\citenamefont {Finklea}\ \emph {et~al.}(1976)\citenamefont
  {Finklea}, \citenamefont {Cathey},\ and\ \citenamefont
  {Amma}}]{Finklea:a12931}%
  \BibitemOpen
  \bibfield  {author} {\bibinfo {author} {\bibfnamefont {S.~L.}\ \bibnamefont
  {Finklea}, \bibfnamefont {III}}, \bibinfo {author} {\bibfnamefont
  {L.}~\bibnamefont {Cathey}}, \ and\ \bibinfo {author} {\bibfnamefont {E.~L.}\
  \bibnamefont {Amma}},\ }\href {\doibase 10.1107/S0567739476001198} {\bibfield
   {journal} {\bibinfo  {journal} {Acta Crystallographica Section A}\ }\textbf
  {\bibinfo {volume} {32}},\ \bibinfo {pages} {529} (\bibinfo {year}
  {1976})}\BibitemShut {NoStop}%
\bibitem [{\citenamefont {Lundqvist}\ and\ \citenamefont
  {Westgren}(1938)}]{10.1002/zaac.19382390110}%
  \BibitemOpen
  \bibfield  {author} {\bibinfo {author} {\bibfnamefont {D.}~\bibnamefont
  {Lundqvist}}\ and\ \bibinfo {author} {\bibfnamefont {A.}~\bibnamefont
  {Westgren}},\ }\href {\doibase 10.1002/zaac.19382390110} {\bibfield
  {journal} {\bibinfo  {journal} {Zeitschrift f{\"u}r anorganische und
  allgemeine Chemie}\ }\textbf {\bibinfo {volume} {239}},\ \bibinfo {pages}
  {85} (\bibinfo {year} {1938})}\BibitemShut {NoStop}%
\bibitem [{\citenamefont {Nowack}\ \emph {et~al.}(1991)\citenamefont {Nowack},
  \citenamefont {Schwarzenbach},\ and\ \citenamefont {Hahn}}]{Nowack:bx0513}%
  \BibitemOpen
  \bibfield  {author} {\bibinfo {author} {\bibfnamefont {E.}~\bibnamefont
  {Nowack}}, \bibinfo {author} {\bibfnamefont {D.}~\bibnamefont
  {Schwarzenbach}}, \ and\ \bibinfo {author} {\bibfnamefont {T.}~\bibnamefont
  {Hahn}},\ }\href {\doibase 10.1107/S0108768191004871} {\bibfield  {journal}
  {\bibinfo  {journal} {Acta Crystallographica Section B}\ }\textbf {\bibinfo
  {volume} {47}},\ \bibinfo {pages} {650} (\bibinfo {year} {1991})}\BibitemShut
  {NoStop}%
\bibitem [{\citenamefont {Streltsov}\ \emph {et~al.}(2017)\citenamefont
  {Streltsov}, \citenamefont {Shorikov}, \citenamefont {Skornyakov},
  \citenamefont {Poteryaev},\ and\ \citenamefont {Khomskii}}]{Streltsov2017}%
  \BibitemOpen
  \bibfield  {author} {\bibinfo {author} {\bibfnamefont {S.~S.}\ \bibnamefont
  {Streltsov}}, \bibinfo {author} {\bibfnamefont {A.~O.}\ \bibnamefont
  {Shorikov}}, \bibinfo {author} {\bibfnamefont {S.~L.}\ \bibnamefont
  {Skornyakov}}, \bibinfo {author} {\bibfnamefont {A.~I.}\ \bibnamefont
  {Poteryaev}}, \ and\ \bibinfo {author} {\bibfnamefont {D.~I.}\ \bibnamefont
  {Khomskii}},\ }\href {\doibase 10.1038/s41598-017-13312-4} {\bibfield
  {journal} {\bibinfo  {journal} {Scientific Reports}\ }\textbf {\bibinfo
  {volume} {7}},\ \bibinfo {pages} {13005} (\bibinfo {year}
  {2017})}\BibitemShut {NoStop}%
\bibitem [{\citenamefont {Jarrett}\ \emph {et~al.}(1968)\citenamefont
  {Jarrett}, \citenamefont {Cloud}, \citenamefont {Bouchard}, \citenamefont
  {Butler}, \citenamefont {Frederick},\ and\ \citenamefont
  {Gillson}}]{jarrett_evidence_1968}%
  \BibitemOpen
  \bibfield  {author} {\bibinfo {author} {\bibfnamefont {H.~S.}\ \bibnamefont
  {Jarrett}}, \bibinfo {author} {\bibfnamefont {W.~H.}\ \bibnamefont {Cloud}},
  \bibinfo {author} {\bibfnamefont {R.~J.}\ \bibnamefont {Bouchard}}, \bibinfo
  {author} {\bibfnamefont {S.~R.}\ \bibnamefont {Butler}}, \bibinfo {author}
  {\bibfnamefont {C.~G.}\ \bibnamefont {Frederick}}, \ and\ \bibinfo {author}
  {\bibfnamefont {J.~L.}\ \bibnamefont {Gillson}},\ }\href {\doibase
  10.1103/PhysRevLett.21.617} {\bibfield  {journal} {\bibinfo  {journal}
  {Physical Review Letters}\ }\textbf {\bibinfo {volume} {21}},\ \bibinfo
  {pages} {617} (\bibinfo {year} {1968})}\BibitemShut {NoStop}%
\bibitem [{\citenamefont {Eyert}\ \emph {et~al.}(1998)\citenamefont {Eyert},
  \citenamefont {H\"ock}, \citenamefont {Fiechter},\ and\ \citenamefont
  {Tributsch}}]{Eyert1998}%
  \BibitemOpen
  \bibfield  {author} {\bibinfo {author} {\bibfnamefont {V.}~\bibnamefont
  {Eyert}}, \bibinfo {author} {\bibfnamefont {K.-H.}\ \bibnamefont {H\"ock}},
  \bibinfo {author} {\bibfnamefont {S.}~\bibnamefont {Fiechter}}, \ and\
  \bibinfo {author} {\bibfnamefont {H.}~\bibnamefont {Tributsch}},\ }\href
  {\doibase 10.1103/PhysRevB.57.6350} {\bibfield  {journal} {\bibinfo
  {journal} {Phys. Rev. B}\ }\textbf {\bibinfo {volume} {57}},\ \bibinfo
  {pages} {6350} (\bibinfo {year} {1998})}\BibitemShut {NoStop}%
\bibitem [{\citenamefont {Folkerts}\ \emph {et~al.}(1987)\citenamefont
  {Folkerts}, \citenamefont {Sawatzky}, \citenamefont {Haas}, \citenamefont
  {de~Groot},\ and\ \citenamefont {Hillebrecht}}]{Folkerts1987}%
  \BibitemOpen
  \bibfield  {author} {\bibinfo {author} {\bibfnamefont {W.}~\bibnamefont
  {Folkerts}}, \bibinfo {author} {\bibfnamefont {G.~A.}\ \bibnamefont
  {Sawatzky}}, \bibinfo {author} {\bibfnamefont {C.}~\bibnamefont {Haas}},
  \bibinfo {author} {\bibfnamefont {R.~A.}\ \bibnamefont {de~Groot}}, \ and\
  \bibinfo {author} {\bibfnamefont {F.~U.}\ \bibnamefont {Hillebrecht}},\
  }\href {\doibase 10.1088/0022-3719/20/26/015} {\bibfield  {journal} {\bibinfo
   {journal} {Journal of Physics C: Solid State Physics}\ }\textbf {\bibinfo
  {volume} {20}},\ \bibinfo {pages} {4135} (\bibinfo {year}
  {1987})}\BibitemShut {NoStop}%
\bibitem [{\citenamefont {Li}\ and\ \citenamefont
  {Xue}(2006)}]{electronegativities}%
  \BibitemOpen
  \bibfield  {author} {\bibinfo {author} {\bibfnamefont {K.}~\bibnamefont
  {Li}}\ and\ \bibinfo {author} {\bibfnamefont {D.}~\bibnamefont {Xue}},\
  }\href {\doibase 10.1021/jp062886k} {\bibfield  {journal} {\bibinfo
  {journal} {The Journal of Physical Chemistry A}\ }\textbf {\bibinfo {volume}
  {110}},\ \bibinfo {pages} {11332} (\bibinfo {year} {2006})}\BibitemShut
  {NoStop}%
\bibitem [{\citenamefont {Vegard}(1921)}]{Vegard1921}%
  \BibitemOpen
  \bibfield  {author} {\bibinfo {author} {\bibfnamefont {L.}~\bibnamefont
  {Vegard}},\ }\href {\doibase 10.1007/BF01349680} {\bibfield  {journal}
  {\bibinfo  {journal} {Zeitschrift f{\"u}r Physik}\ }\textbf {\bibinfo
  {volume} {5}},\ \bibinfo {pages} {17} (\bibinfo {year} {1921})}\BibitemShut
  {NoStop}%
\bibitem [{\citenamefont {Marzari}\ \emph {et~al.}(2012)\citenamefont
  {Marzari}, \citenamefont {Mostofi}, \citenamefont {Yates}, \citenamefont
  {Souza},\ and\ \citenamefont {Vanderbilt}}]{wannierTheory}%
  \BibitemOpen
  \bibfield  {author} {\bibinfo {author} {\bibfnamefont {N.}~\bibnamefont
  {Marzari}}, \bibinfo {author} {\bibfnamefont {A.~A.}\ \bibnamefont
  {Mostofi}}, \bibinfo {author} {\bibfnamefont {J.~R.}\ \bibnamefont {Yates}},
  \bibinfo {author} {\bibfnamefont {I.}~\bibnamefont {Souza}}, \ and\ \bibinfo
  {author} {\bibfnamefont {D.}~\bibnamefont {Vanderbilt}},\ }\href {\doibase
  10.1103/RevModPhys.84.1419} {\bibfield  {journal} {\bibinfo  {journal} {Rev.
  Mod. Phys.}\ }\textbf {\bibinfo {volume} {84}},\ \bibinfo {pages} {1419}
  (\bibinfo {year} {2012})}\BibitemShut {NoStop}%
\bibitem [{\citenamefont {Ramasubramaniam}(2012)}]{ramasubramaniam_large_2012}%
  \BibitemOpen
  \bibfield  {author} {\bibinfo {author} {\bibfnamefont {A.}~\bibnamefont
  {Ramasubramaniam}},\ }\href {\doibase 10.1103/PhysRevB.86.115409} {\bibfield
  {journal} {\bibinfo  {journal} {Phys. Rev. B}\ }\textbf {\bibinfo {volume}
  {86}},\ \bibinfo {pages} {115409} (\bibinfo {year} {2012})}\BibitemShut
  {NoStop}%
\bibitem [{\citenamefont {Shekhar}\ \emph {et~al.}(2015)\citenamefont
  {Shekhar}, \citenamefont {Nayak}, \citenamefont {Sun}, \citenamefont
  {Schmidt}, \citenamefont {Nicklas}, \citenamefont {Leermakers}, \citenamefont
  {Zeitler}, \citenamefont {Skourski}, \citenamefont {Wosnitza}, \citenamefont
  {Liu}, \citenamefont {Chen}, \citenamefont {Schnelle}, \citenamefont
  {Borrmann}, \citenamefont {Grin}, \citenamefont {Felser},\ and\ \citenamefont
  {Yan}}]{shekhar_extremely_2015}%
  \BibitemOpen
  \bibfield  {author} {\bibinfo {author} {\bibfnamefont {C.}~\bibnamefont
  {Shekhar}}, \bibinfo {author} {\bibfnamefont {A.~K.}\ \bibnamefont {Nayak}},
  \bibinfo {author} {\bibfnamefont {Y.}~\bibnamefont {Sun}}, \bibinfo {author}
  {\bibfnamefont {M.}~\bibnamefont {Schmidt}}, \bibinfo {author} {\bibfnamefont
  {M.}~\bibnamefont {Nicklas}}, \bibinfo {author} {\bibfnamefont
  {I.}~\bibnamefont {Leermakers}}, \bibinfo {author} {\bibfnamefont
  {U.}~\bibnamefont {Zeitler}}, \bibinfo {author} {\bibfnamefont
  {Y.}~\bibnamefont {Skourski}}, \bibinfo {author} {\bibfnamefont
  {J.}~\bibnamefont {Wosnitza}}, \bibinfo {author} {\bibfnamefont
  {Z.}~\bibnamefont {Liu}}, \bibinfo {author} {\bibfnamefont {Y.}~\bibnamefont
  {Chen}}, \bibinfo {author} {\bibfnamefont {W.}~\bibnamefont {Schnelle}},
  \bibinfo {author} {\bibfnamefont {H.}~\bibnamefont {Borrmann}}, \bibinfo
  {author} {\bibfnamefont {Y.}~\bibnamefont {Grin}}, \bibinfo {author}
  {\bibfnamefont {C.}~\bibnamefont {Felser}}, \ and\ \bibinfo {author}
  {\bibfnamefont {B.}~\bibnamefont {Yan}},\ }\href {\doibase 10.1038/nphys3372}
  {\bibfield  {journal} {\bibinfo  {journal} {Nature Physics}\ }\textbf
  {\bibinfo {volume} {11}},\ \bibinfo {pages} {645} (\bibinfo {year}
  {2015})}\BibitemShut {NoStop}%
\bibitem [{\citenamefont {Yates}\ \emph {et~al.}(2007)\citenamefont {Yates},
  \citenamefont {Wang}, \citenamefont {Vanderbilt},\ and\ \citenamefont
  {Souza}}]{wanniersFSintegrals}%
  \BibitemOpen
  \bibfield  {author} {\bibinfo {author} {\bibfnamefont {J.~R.}\ \bibnamefont
  {Yates}}, \bibinfo {author} {\bibfnamefont {X.}~\bibnamefont {Wang}},
  \bibinfo {author} {\bibfnamefont {D.}~\bibnamefont {Vanderbilt}}, \ and\
  \bibinfo {author} {\bibfnamefont {I.}~\bibnamefont {Souza}},\ }\href
  {\doibase 10.1103/PhysRevB.75.195121} {\bibfield  {journal} {\bibinfo
  {journal} {Phys. Rev. B}\ }\textbf {\bibinfo {volume} {75}},\ \bibinfo
  {pages} {195121} (\bibinfo {year} {2007})}\BibitemShut {NoStop}%
\bibitem [{\citenamefont {Graser}\ \emph {et~al.}(2009)\citenamefont {Graser},
  \citenamefont {Maier}, \citenamefont {Hirschfeld},\ and\ \citenamefont
  {Scalapino}}]{chi0}%
  \BibitemOpen
  \bibfield  {author} {\bibinfo {author} {\bibfnamefont {S.}~\bibnamefont
  {Graser}}, \bibinfo {author} {\bibfnamefont {T.~A.}\ \bibnamefont {Maier}},
  \bibinfo {author} {\bibfnamefont {P.~J.}\ \bibnamefont {Hirschfeld}}, \ and\
  \bibinfo {author} {\bibfnamefont {D.~J.}\ \bibnamefont {Scalapino}},\ }\href
  {\doibase 10.1088/1367-2630/11/2/025016} {\bibfield  {journal} {\bibinfo
  {journal} {New Journal of Physics}\ }\textbf {\bibinfo {volume} {11}},\
  \bibinfo {pages} {025016} (\bibinfo {year} {2009})}\BibitemShut {NoStop}%
\end{thebibliography}

\end{document}


\section{Supplementary Materials:}

\section{Determination of the Hubbard $U$}

\begin{figure}[htp]
\centering
\includegraphics[width=.4\textwidth]{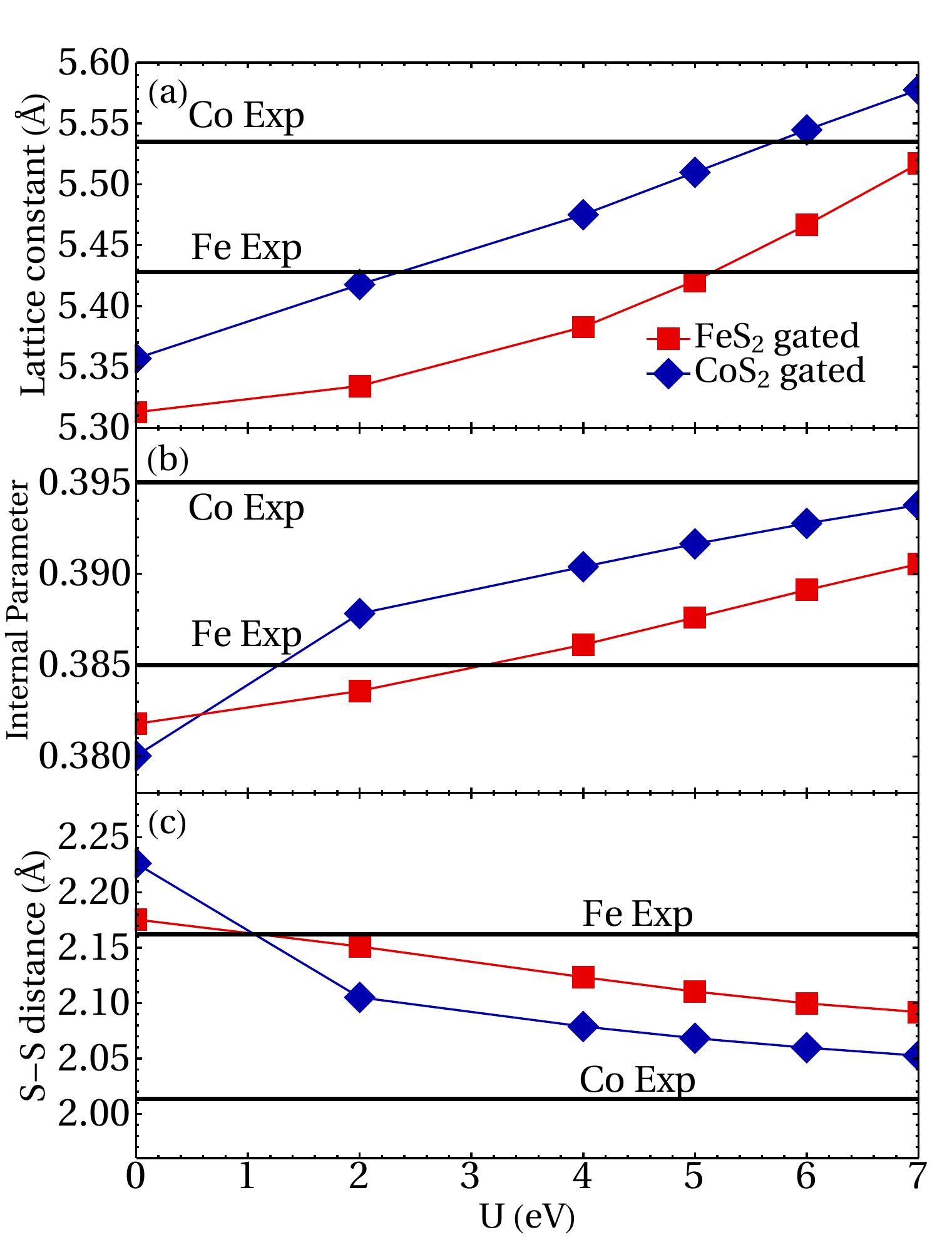}
\caption{The relaxed lattice constant (a), sulfur position parameter $u$ (b), and sulfur-sulfur distance (c) as a function of the Hubbard $U$ in the DFT+U calculations for \chem{CoS_2} in blue diamonds and \chem{FeS_2} in red squares. $U=5$~eV was selected to give reasonable agreement for both \chem{CoS_2} and \chem{FeS_2}.}
\label{fig:Uplots}
\end{figure}

The value of the Hubbard $U$ in our DFT+U calculations was chosen as $U=5$~eV based on the results presented in Figure \ref{fig:Uplots}. This value gives very good agreement with the \chem{FeS_2} lattice constant and $<1\%$ error in the \chem{CoS_2} lattice constant. For the sulfur positions, \chem{FeS_2} has $\sim1-2\%$ error even without U and could be tuned to exact agreement for either $u$ or sulfur-sulfur distance. But \chem{CoS_2} has a worse agreement, and even with U$=7\mathrm{eV}$ the sulfur-sulfur distance is too large. $U=5$~eV was chosen as a compromise that gives similar errors ($\sim2.5\%$) in the sulfur positions in both \chem{FeS_2} and \chem{CoS_2}.

\section{Comparison between tight-binding and DFT band structures}

\begin{figure*}[htp]
\centering
\includegraphics[width=0.65\textwidth]{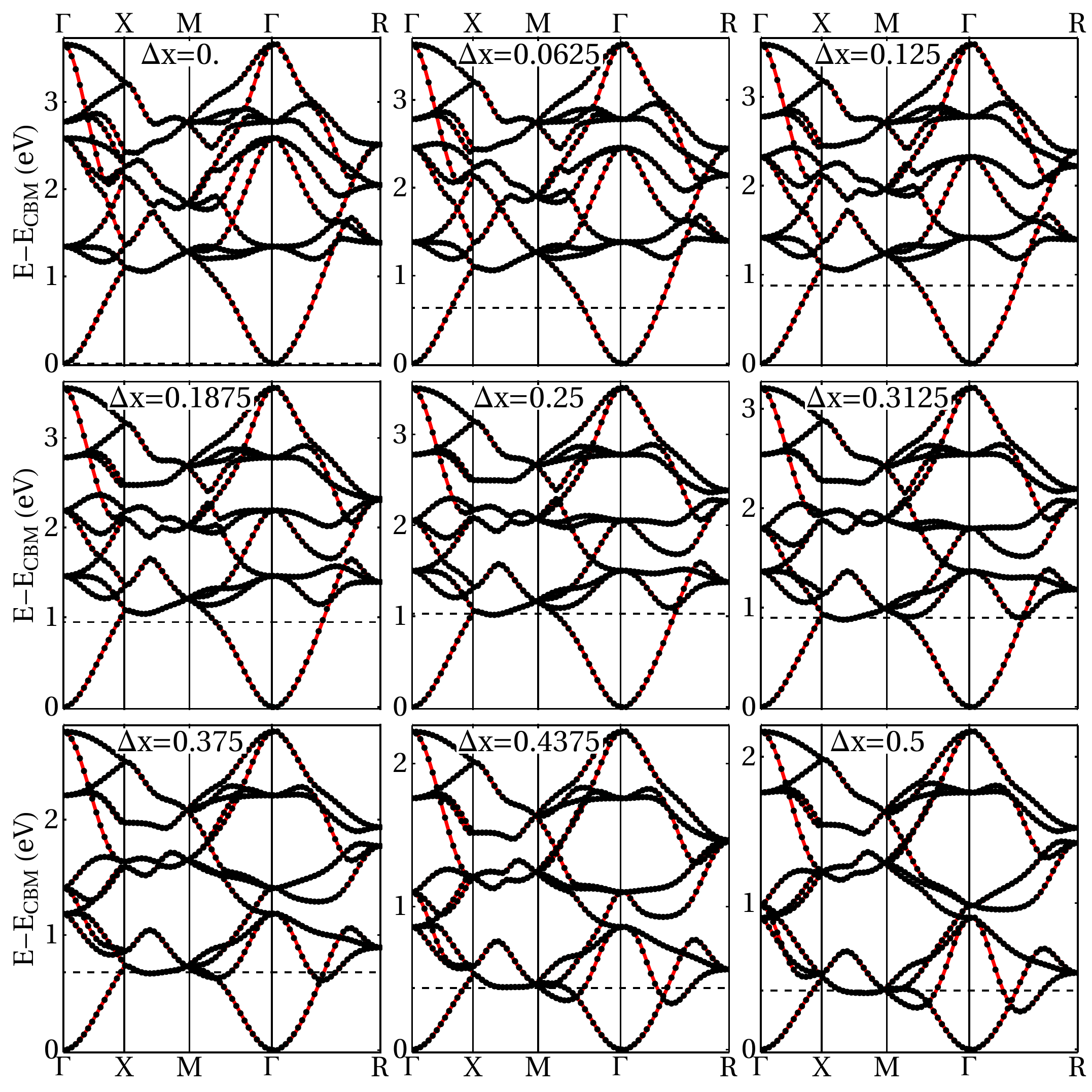}
\caption{Tight-binding band structure of EG \chem{FeS_2} for different added electrons per formula unit (represented here by $\Delta x$ for simplicity). DFT energies are shown by the black dots and energies from the tight-binding model are given by the red lines. The number of added electrons are noted on each panel and the energies are measured from the bottom of the bands, showing a significant change in bandwidth. The Fermi level is marked by a dashed line, which moves from the single sulfur band to the large metallic region at $x\approx0.25$.}
\label{fig:fe-evals}
\end{figure*}

\begin{figure*}[htp]
\centering
\includegraphics[width=0.65\textwidth]{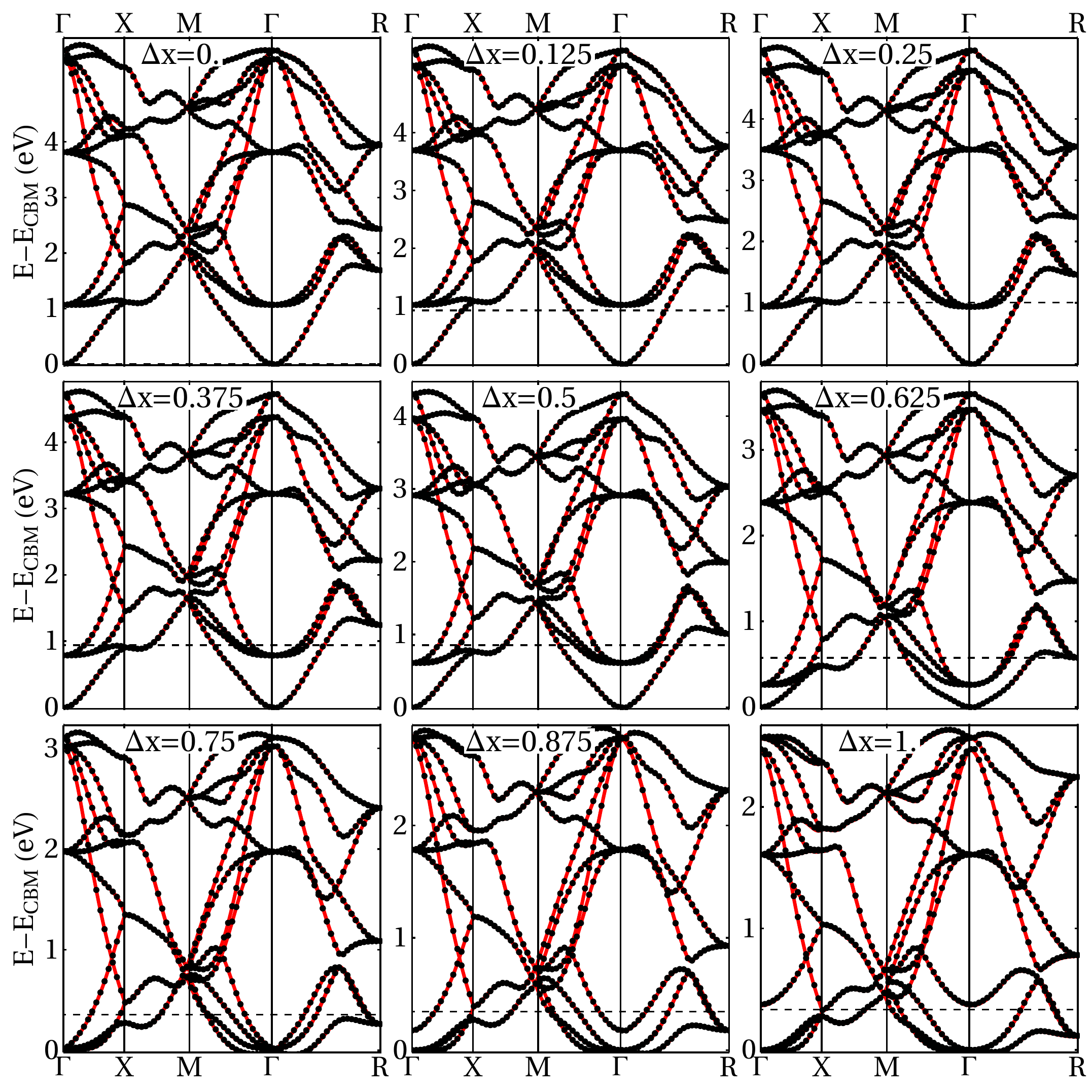}
\caption{Tight-binding band structure of EG \chem{CoS_2} for different added hole per formula unit (represented here by $1-\Delta x$ for simplicity). The notation and color code are the same as in Figure \ref{fig:fe-evals}.}
\label{fig:co-evals}
\end{figure*}

In Figures \ref{fig:fe-evals} and \ref{fig:co-evals} we present the electronic structure of EG \chem{FeS_2} and \chem{CoS_2}, obtained directly from DFT, as well as the Wannier tight binding model. The nearly perfect agreement for all values of added electrons shows the reliability of our Wannier approach. 

